\documentclass[aps,prc,preprint,amsmath,amssymb,showpacs,preprintnumbers,superscriptaddress]{revtex4-2}
\usepackage{graphicx}
\usepackage{dcolumn}
\usepackage{bm}
\usepackage{tabularx}
\usepackage{multirow}
\usepackage{lettrine}
\usepackage{color}
\usepackage{booktabs}

\begin{document}
\title{Revisiting nuclear chirality in $^{128}$Cs with relativistic configuration-interaction density functional theory}

\author{Y. K. Wang}
\email[]{wangyk@buaa.edu.cn}
\affiliation{School of Physics, Beihang University, Beijing 102206, China}

\date{\today}
\begin{abstract}
Nuclear chirality in $^{128}$Cs is revisited within the microscopic relativistic configuration-interaction density functional (ReCD) theory. 
The positive-parity doublet bands are investigated by simultaneously examining their energy spectra, electromagnetic transition probabilities, $g$ factors, spectroscopic quadrupole moments, and underlying angular-momentum geometry. 
Without introducing additional parameters adjusted to the spectroscopic data, the ReCD calculations provide an overall satisfactory description of the available experimental observables. 
In particular, a comprehensive analysis of the available experimental data and the calculated spectroscopic observables indicates qualitative resemblances between the partner bands around $I = 17\hbar$.
A microscopic analysis of the angular-momentum geometry through \textit{azimuthal plots} reveals a distinct evolution of the rotational mode with increasing spin: chiral vibration at $I < 17\hbar$, static chirality at $I = 17\hbar$, and a transition toward planar rotation at higher spins. 
These results suggest that static chiral geometry in $^{128}$Cs is confined to a narrow spin region around $I = 17\hbar$ within the present ReCD calculations.
\end{abstract}
\maketitle
\date{today}

\section{Introduction}
Nuclear chiral rotation is a fascinating rotational mode that exhibits distinctive features in the generation of angular momenta, energy spectra, and electromagnetic transitions, in contrast to conventional collective rotation characterized by substantial quadrupole deformation and strong $E2$ transitions. 
In a chirally rotating nucleus, the angular momenta carried by the high-$j$ proton (neutron) particle, high-$j$ neutron (proton) hole, and the triaxially deformed core tend to align along the short, long, and intermediate principle axes, respectively, forming an aplanar angular-momentum geometry~\cite{Frauendorf:1997mux}. 
Such a configuration can occur in two energetically equivalent arrangements with opposite handedness, corresponding to left- and right-handed systems, and thereby gives rise to spontaneous chiral symmetry breaking in the intrinsic frame. 
In the laboratory frame, quantum tunneling and mixing between the two handed configurations restore the chiral symmetry, leading to the characteristic manifestation of a pair of nearly degenerate $\Delta I=1$ rotational bands with the same parity and similar spectroscopic properties, commonly referred to as chiral doublet bands. 
Since the emergence of nuclear chirality requires a triaxial nuclear shape, the observation and identification of chiral doublet bands provide an important spectroscopic signature of nuclear triaxiality. 
Owing to its fundamental interest as a manifestation of chirality at the subatomic scale and its intimate connection with nuclear triaxial deformation, nuclear chiral rotation has attracted extensive experimental~\cite{Wang:2011un,Liu:2016ddt,Vaman:2004zz,Joshi:2004yyg,Tonev:2014iha,Lieder:2014ria,Rather:2013zos,Kuti:2014zla,Starosta:2001zz,Zhu:2003yu,Tonev:2006zz,Petrache:2006zz,PhysRevLett.97.172501,Tonev:2007zz,Mukhopadhyay:2007zx,Ayangeakaa:2013ck,Petrache:2018tla,Lv:2019ewz,Balabanski:2004hs,Lawrie:2008zzb} and theoretical~\cite{Koike:2004zz,Zhang:2007ge,Qi:2008zp,Lawrie:2010zzb,Chen:2018sys,Wang:2019ehn,Budaca:2019dnv,Dimitrov:2000np,Olbratowski:2002zb,Zhao:2017opp,Chen:2012vk,Ren:2022fau,Li:2024fmh,Chen:2017cwl,Wang:2019inb,Wang:2023qll} investigations over the past three decades; see Refs.~\cite{Meng:2010kj,Meng:2016jai,Raduta:2016hjl,Chen:2020zsy,Xiong:2018tgf} for recent reviews and data compilations.

The emergence of nuclear chirality generally requires high-$j$ valence particles and holes coupled to a triaxially deformed core. 
Consequently, candidate chiral nuclei are predominantly found in several specific mass regions of the nuclear chart, around $A \sim 80$~\cite{Wang:2011un,Liu:2016ddt}, 100~\cite{Vaman:2004zz,Joshi:2004yyg,Tonev:2014iha,Lieder:2014ria,Rather:2013zos,Kuti:2014zla}, 130~\cite{Starosta:2001zz,Zhu:2003yu,Tonev:2006zz,Petrache:2006zz,PhysRevLett.97.172501,Tonev:2007zz,Mukhopadhyay:2007zx,Ayangeakaa:2013ck,Petrache:2018tla,Lv:2019ewz}, and 190~\cite{Balabanski:2004hs,Lawrie:2008zzb}. 
Among the candidates identified so far, $^{128}$Cs has attracted particular attention. 
It is one of the few chiral candidates for which lifetime measurements have been performed for both partner bands, providing experimental $B(E2)$ and $B(M1)$ values that offer crucial information for assessing their chiral nature. 
In particular, the similarity of the electromagnetic transition probabilities between the two positive-parity bands led to $^{128}$Cs being regarded as one of the best examples of chiral symmetry breaking in atomic nuclei~\cite{PhysRevLett.97.172501}.

Despite this compelling experimental evidence, however, the microscopic nature and the degree of chirality in $^{128}$Cs remain unsettled. 
Beyond the energy spectra and electromagnetic transition probabilities conventionally used to identify chiral doublet bands, the $g$ factor of the bandhead state has also been measured~\cite{Grodner:2018jgx}, providing an additional and independent probe of its underlying angular-momentum structure. 
A quantitative analysis of the measured $g$ factor within the particle-rotor model (PRM) indicated a nonchiral character of the bandhead state and suggested that static chirality in $^{128}$Cs, if realized, emerges only above a critical angular momentum~\cite{Grodner:2018jgx,Grodner:2022cnf}. 
This conclusion is consistent with the angular-momentum-projected (AMP) calculations based on a pairing-plus-quadrupole Hamiltonian in Ref.~\cite{Chen:2017cwl}, where an explicit analysis of the angular-momentum geometry suggested that static chiral rotation develops only around $I=14\hbar$, while the lower-spin states are dominated by non-chiral planar rotation, namely, chiral vibration. 
More strikingly, calculations within the nonadiabatic quasiparticle approach (NQA) in Ref.~\cite{Siwach:2021xgj} did not support the formation of static chirality in $^{128}$Cs even at higher spins. 
Therefore, although $^{128}$Cs was once regarded as a good example of nuclear chirality, the emergence and development of static chirality still deserve further clarification.

It should be noted that previous theoretical studies of $^{128}$Cs, including those based on the PRM~\cite{Grodner:2018jgx,Grodner:2022cnf}, AMP~\cite{Chen:2017cwl}, and NQA~\cite{Siwach:2021xgj}, rely either on a phenomenological description of the triaxial core or on a phenomenological pairing-plus-quadrupole Hamiltonian. 
Consequently, these calculations generally involve model parameters that need to be determined, at least in part, by experimental data. 
In the present work, we revisit the positive-parity candidate chiral doublet bands in $^{128}$Cs within the framework of relativistic configuration-interaction density functional (ReCD) theory.
The ReCD theory is a microscopic many-body framework that combines the advantages of nuclear relativistic density functional theory (DFT)~\cite{meng2016relativistic} and the large-scale configuration-interaction shell model~\cite{Caurier:2004gf}.
It has been successfully applied to a variety of nuclear ground-state~\cite{Qu:2026vsb}, excited-state~\cite{Zhao:2016umr,Wang:2022wxl,Wang:2023qll,Qu:2025qvy}, and weak-decay properties~\cite{Wang:2024zkl,Wang:2023hkc,Wang:2026kje}. 
The calculation starts from a self-consistent relativistic DFT solution, in which nuclear deformation and pairing correlations are treated simultaneously. 
A configuration space is subsequently constructed by generating quasiparticle excitations on top of the self-consistent reference state. 
Three-dimensional angular momentum projection is then performed for all intrinsic configurations, yielding a set of projected basis states with good angular momentum. 
Importantly, angular momentum projection not only restores the rotational symmetry broken at the mean-field level but also provides the transformation from the intrinsic description to the laboratory frame. 
This enables the ReCD framework to incorporate the quantum mixing between left- and right-handed intrinsic configurations that is essential for describing chiral doublet bands. 
Finally, the correlated many-body wave functions in the laboratory frame are obtained by diagonalizing the many-body Hamiltonian in the space spanned by the projected basis states.
To provide a comprehensive assessment of the chiral nature of the positive-parity doublet bands in $^{128}$Cs, we calculate their energy spectra and electromagnetic transition probabilities, as well as the $g$ factors and spectroscopic quadrupole moments, thereby confronting the theory with a broad range of available experimental observables. 
Furthermore, the underlying angular-momentum geometry and its evolution with spin are investigated using the \textit{azimuthal plot} introduced in Ref.~\cite{Chen:2017cwl}, providing a direct microscopic characterization of the development of chiral rotation in $^{128}$Cs.

\section{Theoretical framework}
In the ReCD theory, the nuclear wave functions with good angular momentum $I$ and its projection $M$ onto the laboratory $z$ axis are expressed as~\cite{Wang:2024zkl,Wang:2023hkc,Wang:2026kje}
\begin{align}\label{eq:Psi}
|\Psi^{IM}_{\sigma}\rangle = \sum_{K\kappa} F^{I\sigma}_{K\kappa} \hat{P}^{I}_{MK} |\Phi_{\kappa}\rangle,
\end{align}
where $\sigma$ distinguishes different nuclear states with the same angular momentum $I$, and $\hat{P}^{I}_{MK}$ denotes the three-dimensional angular-momentum-projection operator~\cite{Ring2004}. 
The intrinsic configurations $|\Phi_{\kappa}\rangle$ include the mean-field ground state as well as two- and four-quasiparticle (qp) excited states. 
For the odd-odd nucleus $^{128}$Cs considered in the present work, the ReCD configuration space is constructed as
\begin{align}\label{eq:config.space}
\big\{
\hat{\beta}^{\dagger}_{\pi_0}\hat{\beta}^{\dagger}_{\nu_0}|\Phi_0\rangle,
\hat{\beta}^{\dagger}_{\pi_i}\hat{\beta}^{\dagger}_{\nu_j}|\Phi_0\rangle,
\hat{\beta}^{\dagger}_{\pi_i}\hat{\beta}^{\dagger}_{\nu_j}\hat{\beta}^{\dagger}_{\pi_k}\hat{\beta}^{\dagger}_{\pi_l}|\Phi_0\rangle,
\hat{\beta}^{\dagger}_{\pi_i}\hat{\beta}^{\dagger}_{\nu_j}\hat{\beta}^{\dagger}_{\nu_k}\hat{\beta}^{\dagger}_{\nu_l}|\Phi_0\rangle\big\},
\end{align}
where $\hat{\beta}^{\dagger}_{\pi}$ and $\hat{\beta}^{\dagger}_{\nu}$ are the qp creation operators for protons and neutrons, respectively. 
All intrinsic configurations in Eq.~\eqref{eq:config.space} are obtained by iteratively solving the triaxial relativistic Hartree-Bogoliubov (TRHB) equation~\cite{meng2016relativistic}. 
In particular, $\hat{\beta}^{\dagger}_{\pi_0}\hat{\beta}^{\dagger}_{\nu_0}|\Phi_0\rangle
\equiv |\Phi_{\pi_0\nu_0}\rangle$ denotes the mean-field ground state of the odd-odd system. Here, $\pi_0$ and $\nu_0$ label the proton and neutron qp orbitals with the lowest qp excitation energies, respectively, which are blocked during the iterative calculation of the TRHB equation to ensure the correct number parity of the odd-odd system~\cite{Ring2004}.

Applying the angular-momentum-projection operator $\hat{P}^{I}_{MK}$ to the intrinsic configurations $|\Phi_{\kappa}\rangle$ generates the projected basis states $\{\hat{P}^{I}_{MK}|\Phi_{\kappa}\rangle\}$ with good angular momentum. 
The many-body Hamiltonian $\hat{H}$ is then diagonalized in the Hilbert space spanned by these projected states. 
This leads to the Hill-Wheeler equation
\begin{align}\label{eq:Hill.Wheeler}
\sum_{K\kappa} \left( H^{I}_{K'\kappa'K\kappa} - E^{I\sigma} N^{I}_{K'\kappa'K\kappa} \right) F^{I\sigma}_{K\kappa} = 0,
\end{align}
where the Hamiltonian and norm kernels are defined, respectively, as
\begin{align}\label{eq:kernels}
H^{I}_{K'\kappa'K\kappa}
=
\langle\Phi_{\kappa'}|
\hat{H}\hat{P}^{I}_{K'K}
|\Phi_{\kappa}\rangle,
N^{I}_{K'\kappa'K\kappa}
=
\langle\Phi_{\kappa'}|
\hat{P}^{I}_{K'K}
|\Phi_{\kappa}\rangle.
\end{align}
The many-body Hamiltonian $\hat{H}$ is derived from the relativistic Lagrangian density through a Legendre transformation~\cite{meng2016relativistic}. 
Both the Hamiltonian and norm kernels are evaluated using the Pfaffian algorithms developed in Refs.~\cite{Hu:2013dha,Carlsson:2020fiv}. 
Solving Eq.~\eqref{eq:Hill.Wheeler} yields the energies $E^{I\sigma}$ and expansion coefficients $F^{I\sigma}_{K\kappa}$, and thereby determines the correlated many-body wave functions $|\Psi^{IM}_{\sigma}\rangle$ in Eq.~\eqref{eq:Psi}.

With the resulting many-body wave functions, the reduced electromagnetic transition probabilities are calculated as
\begin{align}\label{eq:E2.M1}
B(E\ {\rm or}\ M\lambda;I\sigma\rightarrow I'\sigma')
=
\frac{1}{2I+1}
\left|
\langle
\Psi^{I'\sigma'}
\Vert
\hat{T}_{\lambda}
\Vert
\Psi^{I\sigma}
\rangle
\right|^2,
\end{align}
where $\hat{T}_{\lambda\mu}$ denotes an electromagnetic multipole operator of rank $\lambda$. For the $E2$ transition, the electric quadrupole operator is given by~\cite{Ring2004}
\begin{align}
\hat{T}_{2\mu}
\equiv
\hat{Q}_{2\mu}
=
e\sum_{i=1}^{A}
\left(\frac{1}{2}-t_3^{(i)}\right)
r_i^2Y_{2\mu}(\theta_i,\phi_i),
\end{align}
where $t_3^{(i)} = -1/2$ for protons and $+1/2$ for neutrons. 
For the $M1$ transition, the magnetic dipole operator reads~\cite{Ring2004}
\begin{align}
\hat{T}_{1\mu}
\equiv
\hat{M}_{1\mu}
=
\mu_N
\sum_{i=1}^{A}
\left\{
g_s\hat{\bm{s}}_i
+
g_l\hat{\bm{l}}_i
\right\}
\cdot
\left[
\bm{\nabla}
rY_{1\mu}(\theta,\phi)
\right]_{\bm r=\bm r_i},
\end{align}
where $\mu_N$ is the nuclear magneton, while $g_l$ and $g_s$ denote the orbital and spin $g$ factors of the nucleons, respectively. 
The spectroscopic quadrupole moment $Q_s^{I\sigma}$ and the $g^{I\sigma}$ factor of a nuclear state can likewise be evaluated as
\begin{align}
Q_s^{I\sigma}
=
\sqrt{\frac{16\pi}{5}}
\langle
\Psi^{II}_{\sigma}
|
\hat{Q}_{20}
|
\Psi^{II}_{\sigma}
\rangle,\quad
g^{I\sigma}
=
\sqrt{\frac{4\pi}{3}}
\frac{
\langle
\Psi^{II}_{\sigma}
|
\hat{M}_{10}
|
\Psi^{II}_{\sigma}
\rangle
}{I \mu_N}.
\end{align}

To elucidate the angular-momentum geometry underlying chiral rotation and its evolution with spin, it is desirable to extract the orientation probability distribution of the total angular momentum in the intrinsic frame. 
Such an extraction is nontrivial because the projected states $\{\hat{P}^{I}_{MK}|\Phi_{\kappa}\rangle\}$ entering the expansion of $|\Psi^{IM}_{\sigma}\rangle$ are defined in the laboratory frame and constitute a nonorthogonal basis. 
This difficulty can be circumvented by introducing the \textit{azimuthal plot} proposed in Ref.~\cite{Chen:2017cwl}, which provides the probability distribution for the orientation of the total angular momentum on the intrinsic $(\theta,\phi)$ plane. 
Following Ref.~\cite{Chen:2017cwl}, the \textit{azimuthal plot} is defined as
\begin{align}\label{eq:A-plot}
\mathcal{P}^{I\sigma}(\theta,\phi)
=
\sum_{\kappa}
\int_{0}^{2\pi}
d\psi'
\left|
W^{I\sigma}_{\kappa}
(\psi',\theta,\pi-\phi)
\right|^2 ,
\end{align}
where
\begin{align}
W^{I\sigma}_{\kappa}(\psi',\theta,\pi-\phi)
=
\sqrt{\frac{2I+1}{8\pi^2}}
\sum_K
g^{I\sigma}_{K\kappa}
D^{I*}_{IK}
(\psi',\theta,\pi-\phi).
\end{align}
Here, $g^{I\sigma}_{K\kappa}$ denotes the collective wave function and is related to the expansion coefficients $F^{I\sigma}_{K\kappa}$ through
\begin{align}
g^{I\sigma}_{K\kappa}
=
\sum_{K'\kappa'}
(N^I)^{1/2}_{K\kappa K'\kappa'}
F^{I\sigma}_{K'\kappa'},
\end{align}
where $(N^I)^{1/2}_{K\kappa K'\kappa'}$ is the matrix element of the square root of the norm matrix defined in Eq.~\eqref{eq:kernels}. 
The polar angle $\theta$ specifies the angle between the total angular momentum $I$ and the intrinsic long ($l$) axis, whereas the azimuthal angle $\phi$ specifies the angle between the projection of $I$ onto the intrinsic intermediate-short ($i$-$s$) plane and the intermediate ($i$) axis. 
The angles $(\theta,\phi)$ are related to the Euler angles $\Omega=(\psi',\theta',\phi')$ through $\theta=\theta'$ and $\phi=\pi-\phi'$, following the convention adopted in Ref.~\cite{Chen:2017cwl}.

\section{Results and discussion}
In the following, the positive-parity chiral doublet bands in $^{128}$Cs, denoted as the Yrast and Side bands, are investigated within the ReCD framework. 
The relativistic density functional PC-PK1~\cite{Zhao:2010hi} is employed to construct the many-body Hamiltonian $\hat{H}$ and the corresponding TRHB equation. 
Pairing correlations are treated using a finite-range separable pairing force with a strength of $G=728~\text{MeV}\cdot\text{fm}^{3}$~\cite{Tian:2009zzh}. 
The TRHB equation is solved in a three-dimensional harmonic-oscillator basis containing 10 major shells, which has been shown to provide sufficient numerical accuracy for nuclei in the $A\sim130$ mass region~\cite{Zhao:2015bua,Wang:2022xls,Guo:2019gtc}.
In solving the TRHB equation for the odd-odd nucleus $^{128}$Cs, the proton and neutron $h_{11/2}$ qp orbitals with the lowest qp excitation energies are blocked, in accordance with the $\nu h_{11/2}\otimes\pi h_{11/2}$ configuration assigned to the observed chiral doublet bands~\cite{Grodner:2018jgx}. 
By examining the potential energy surface, the equilibrium deformation that minimizes the bandhead energy is determined to be $(\beta,\gamma)=(0.23,26^\circ)$.
The pronounced triaxial deformation with $\gamma=26^\circ$ provides a favorable intrinsic geometry for the development of chiral rotation in $^{128}$Cs.
As in our previous ReCD studies~\cite{Zhao:2016umr,Wang:2022wxl,Wang:2023qll,Qu:2025qvy,Wang:2024zkl,Wang:2023hkc,Wang:2026kje}, the dimension of the configuration space in Eq.~\eqref{eq:config.space} is controlled by imposing a cutoff $E_{\mathrm{cut}}$ on the qp excitation energies. 
In the present calculation, $E_{\mathrm{cut}}=4.5$ MeV is adopted. 
We have verified that increasing the cutoff further produces negligible changes in the observables discussed below, demonstrating that the calculated results are well converged with respect to the configuration-space truncation. 
Apart from the parameters already fixed in the PC-PK1 functional and the pairing interaction, no additional parameters are introduced or adjusted to the spectroscopic data of $^{128}$Cs.

\begin{figure}[htbp]
  \centering
  \includegraphics[width=0.5\textwidth]{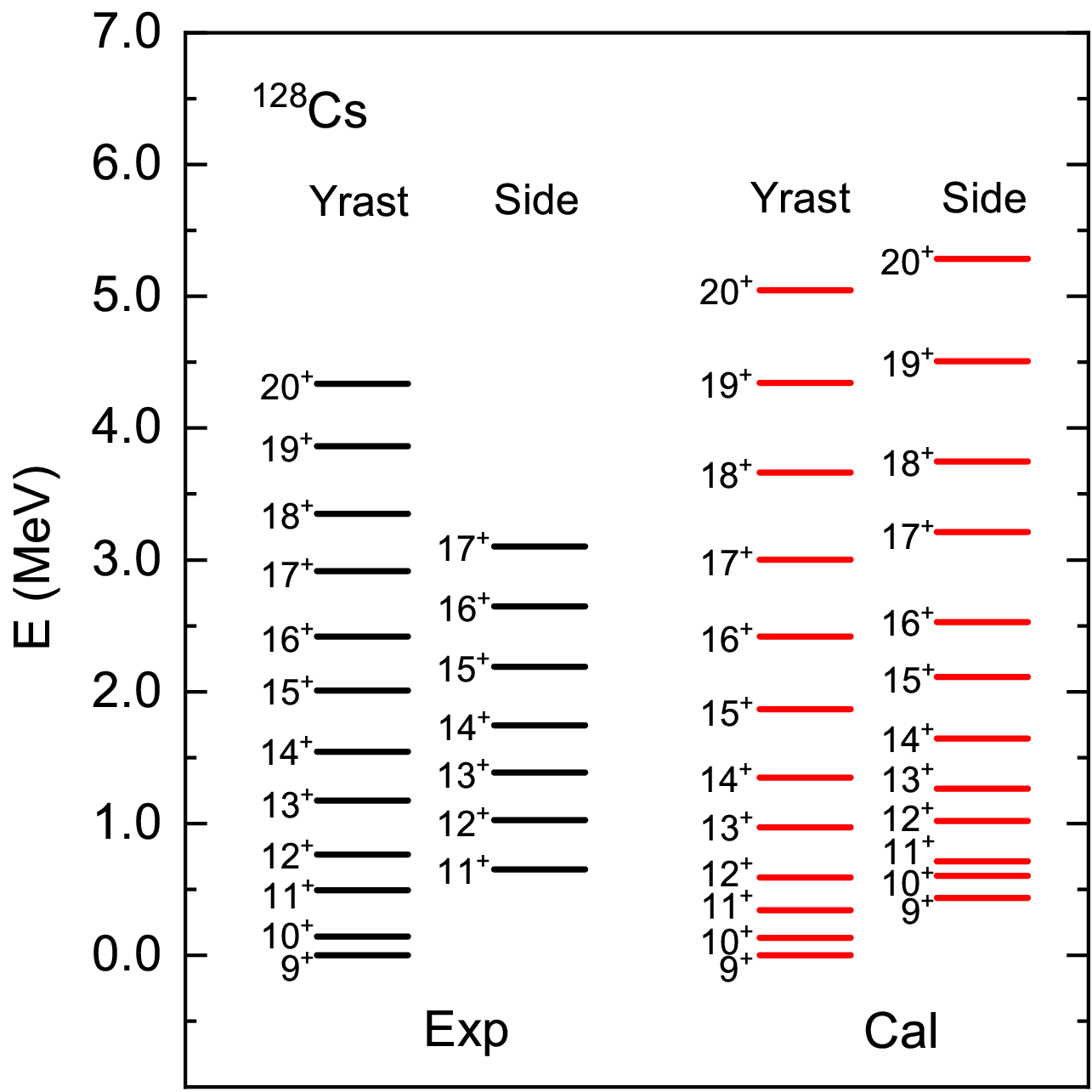}
  \caption{(Color online) Energy spectra of the positive-parity chiral doublet bands in $^{128}$Cs obtained within the ReCD theory in comparison with the experimental data~\cite{Grodner:2018jgx}. 
  The energies are given relative to the $I=9^+$ bandhead state of the Yrast band.}
  \label{Fig:spectrum}
\end{figure}

In Fig.~\ref{Fig:spectrum}, the energy spectra of the Yrast and Side bands calculated within the ReCD theory are compared with the experimental data~\cite{Grodner:2018jgx}. 
Overall, the calculated spectra reproduce the experimental data satisfactorily, including the characteristic near degeneracy between the Yrast and Side bands. 
At higher spins, however, the calculated excitation energies gradually deviate from the experimental values and become systematically overestimated for $I\geq18\hbar$. 
For example, the calculated excitation energy of the $I=20^+$ state is approximately $0.7$ MeV higher than the experimental value. 
Since this discrepancy emerges predominantly in the high-spin region, where higher-order qp excitations are expected to become increasingly relevant, it may be attributed, at least partly, to the present truncation of the ReCD configuration space at the four-qp level. 
Extending the configuration space to include higher-order qp configurations may therefore improve the description of the high-spin states.

\begin{figure}[htbp]
  \centering
  \includegraphics[width=0.5\textwidth]{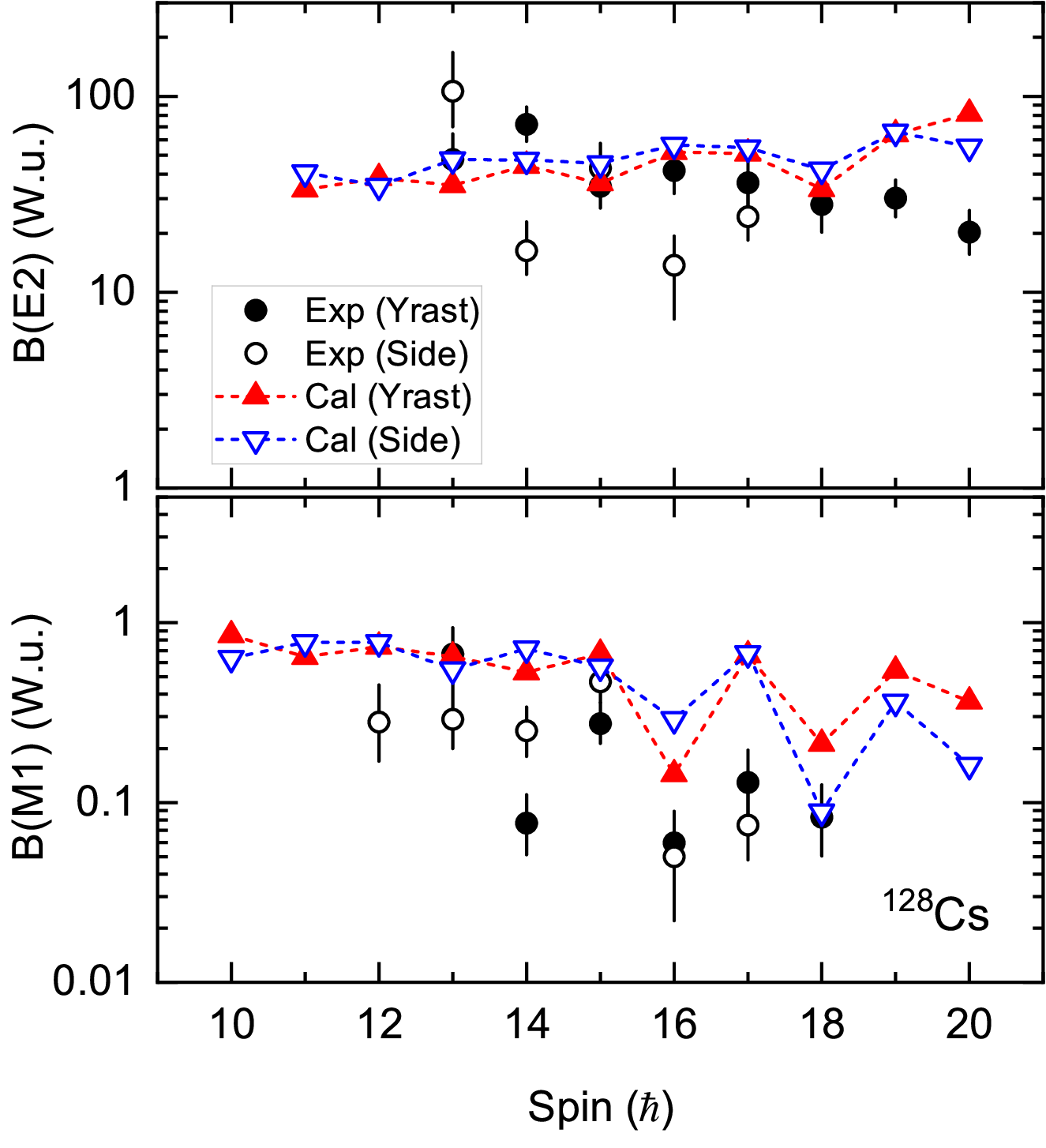}
  \caption{(Color online) $E2$ and $M1$ transition probabilities of the Yrast and Side bands calculated by the ReCD theory, in comparison with the experimental data~\cite{Grodner:2018jgx}.}
  \label{Fig:transition}
\end{figure}

As pointed out in a previous study~\cite{Petrache:2006zz}, the near degeneracy of the partner bands alone is insufficient to establish their chiral nature. 
Electromagnetic transition probabilities, in particular the $B(E2)$ and $B(M1)$ values, provide important complementary information for assessing the emergence and development of chiral rotation. 
Figure~\ref{Fig:transition} presents the $B(E2)$ and $B(M1)$ values obtained from the ReCD calculations in comparison with the available experimental data. 
Owing to the large single-particle space employed in the ReCD framework, neither effective charges nor effective orbital and spin $g$ factors are introduced in the present calculations.

For the $B(E2)$ values, the ReCD calculations reproduce the experimental data well up to $I=18\hbar$. 
At higher spins, the calculated values become larger than the experimental ones, similar to the increasing deviation observed for the excitation energies in Fig.~\ref{Fig:spectrum}. 
As discussed above, this discrepancy may be partly associated with the truncation of the present ReCD configuration space at the four-qp level, since higher-order qp configurations are expected to become increasingly relevant at high spins.
The calculated $B(E2)$ values of the Yrast and Side bands remain remarkably similar over the entire spin range considered, consistent with one of the characteristic expectations for chiral partner bands.

For the $B(M1)$ values, the ReCD calculations reproduce the overall magnitude and spin dependence of the experimental data reasonably well. 
A systematic overestimation of the measured values is nevertheless observed, including in the low-spin region. 
Since both the excitation energies and $B(E2)$ values are well reproduced in the low-spin region, the systematic deviation in $B(M1)$ suggests that effects beyond the one-body $M1$ operator adopted in the present calculation may play an important role. 
Indeed, previous studies have demonstrated that meson-exchange currents can provide non-negligible corrections to magnetic observables~\cite{Harting:1981cc,Kohno:1982zz}. 
Their inclusion in the $M1$ operator may therefore modify the calculated $B(M1)$ strengths and potentially improve the agreement with experiment. 
Nevertheless, considering that no effective $g$ factors or additional parameters are introduced in this work, the overall agreement achieved by the present microscopic calculation is satisfactory.

Of particular interest is the pronounced decrease of the $B(M1)$ strength at $I=16\hbar$, which is well captured by the ReCD calculation. 
Moreover, the calculated $B(M1)$ values of the Yrast and Side bands exhibit their closest similarity around $I=17\hbar$, in agreement with the experimental trend. 
At higher spins, particularly for $I\geq18\hbar$, the similarity between the $B(M1)$ values of the two bands becomes less pronounced. 
As will be discussed below in connection with the angular-momentum geometry, this spin-dependent behavior provides useful information on the evolution of chirality in $^{128}$Cs.

\begin{figure}[htbp]
  \centering
  \includegraphics[width=0.5\textwidth]{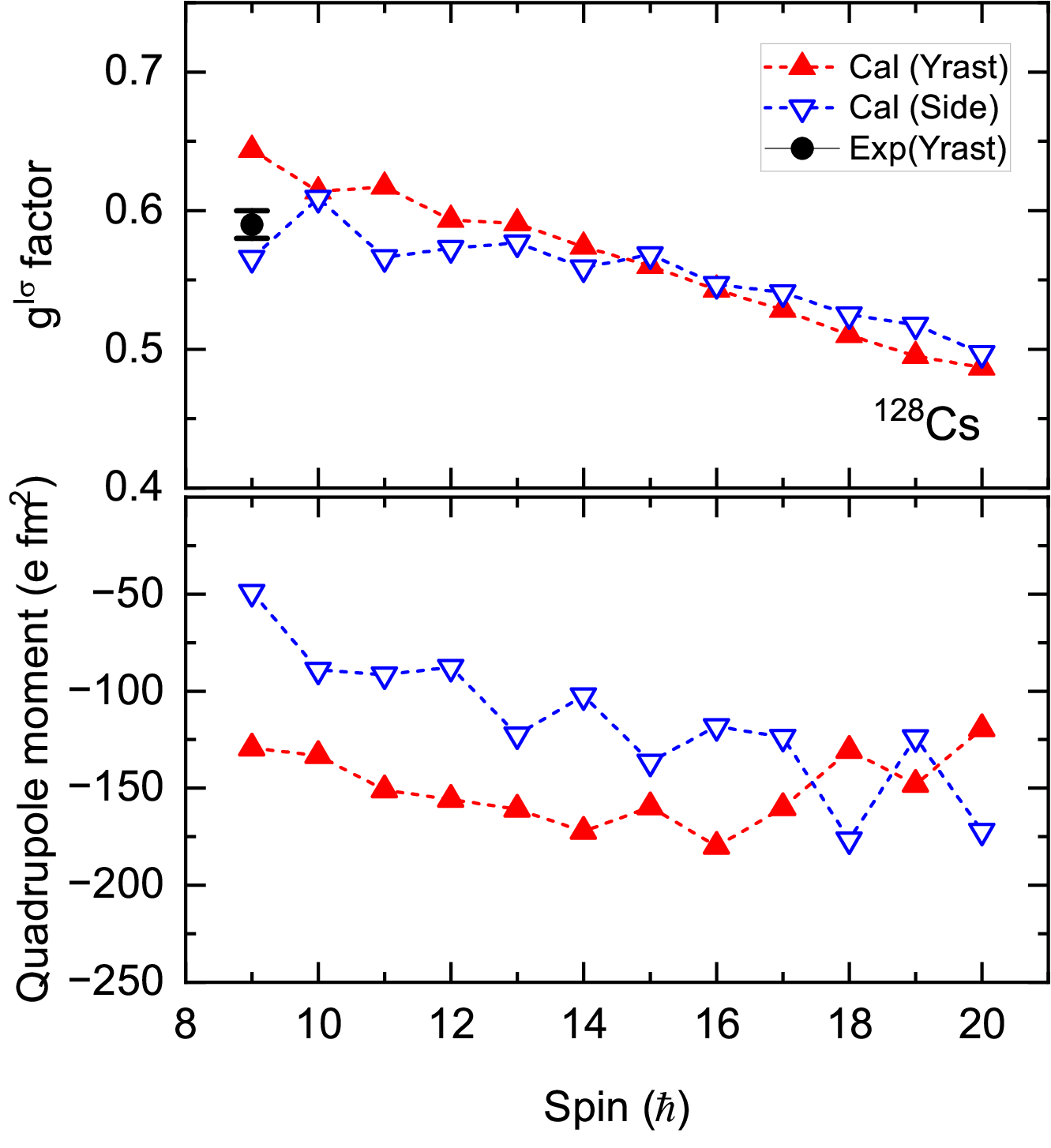}
  \caption{(Color online) $g^{I\sigma}$ and spectroscopic quadrupole moment $Q^{I\sigma}_s$ of the Yrast and Side bands calculated by the ReCD theory, in comparison with the available experimental data~\cite{Grodner:2022cnf}.}
  \label{Fig:gfact.Q2}
\end{figure}

Recently, the $g$ factors and spectroscopic quadrupole moments $Q_s$ of chiral doublet bands have attracted increasing attention as additional observables for probing nuclear chirality~\cite{Grodner:2022cnf,Chen:2020kwc,Hu:2024vwn,Chen:2024psf}. 
In contrast to the energy spectra and electromagnetic transition probabilities, these quantities provide complementary information on the magnetic and quadrupole properties of individual nuclear states and thus offer additional constraints on the similarity between the chiral partner bands. 
Figure~\ref{Fig:gfact.Q2} presents the $g$ factors and spectroscopic quadrupole moments calculated within the ReCD theory. 
Experimentally, only the $g$ factor of the $I=9^+$ bandhead state of the Yrast band is currently available.

For the $g$ factors, the calculated values for the Yrast and Side bands exhibit similar overall trends and decrease gradually with increasing spin. 
At $I=9\hbar$, the calculated $g$ factor of the Yrast band reproduces the experimental value satisfactorily. 
A closer inspection reveals that the difference between the $g$ factors of the two bands generally decreases with increasing spin up to $I\approx16\hbar$, indicating an increasing similarity of their magnetic properties. 
Beyond this spin region, the two bands begin to exhibit more noticeable differences.

A somewhat different behavior is found for the spectroscopic quadrupole moments. 
In contrast to the $g$ factors, a close similarity between the $Q_s$ values of the two partner bands is not systematically established over the entire spin range. 
Nevertheless, their differences become relatively small at several spins, particularly around $I=13\hbar$, $15\hbar$, and $17\hbar$. 
Combining these results with the $B(E2)$ and $B(M1)$ values shown in Fig.~\ref{Fig:transition} reveals qualitative similarities between the partner bands around $I=17\hbar$. 
This observation suggests that the chiral characteristics of the two bands may be most pronounced in this spin region.
It should be emphasized, however, that the ``similarity'' between the corresponding observables of the partner bands remains, to some extent, a qualitative criterion and does not by itself provide a quantitative measure of the degree of chirality. 
A more direct characterization therefore requires an examination of the underlying angular-momentum geometry, which will be discussed below.

\begin{figure}[htbp]
  \centering
  \includegraphics[width=0.8\textwidth]{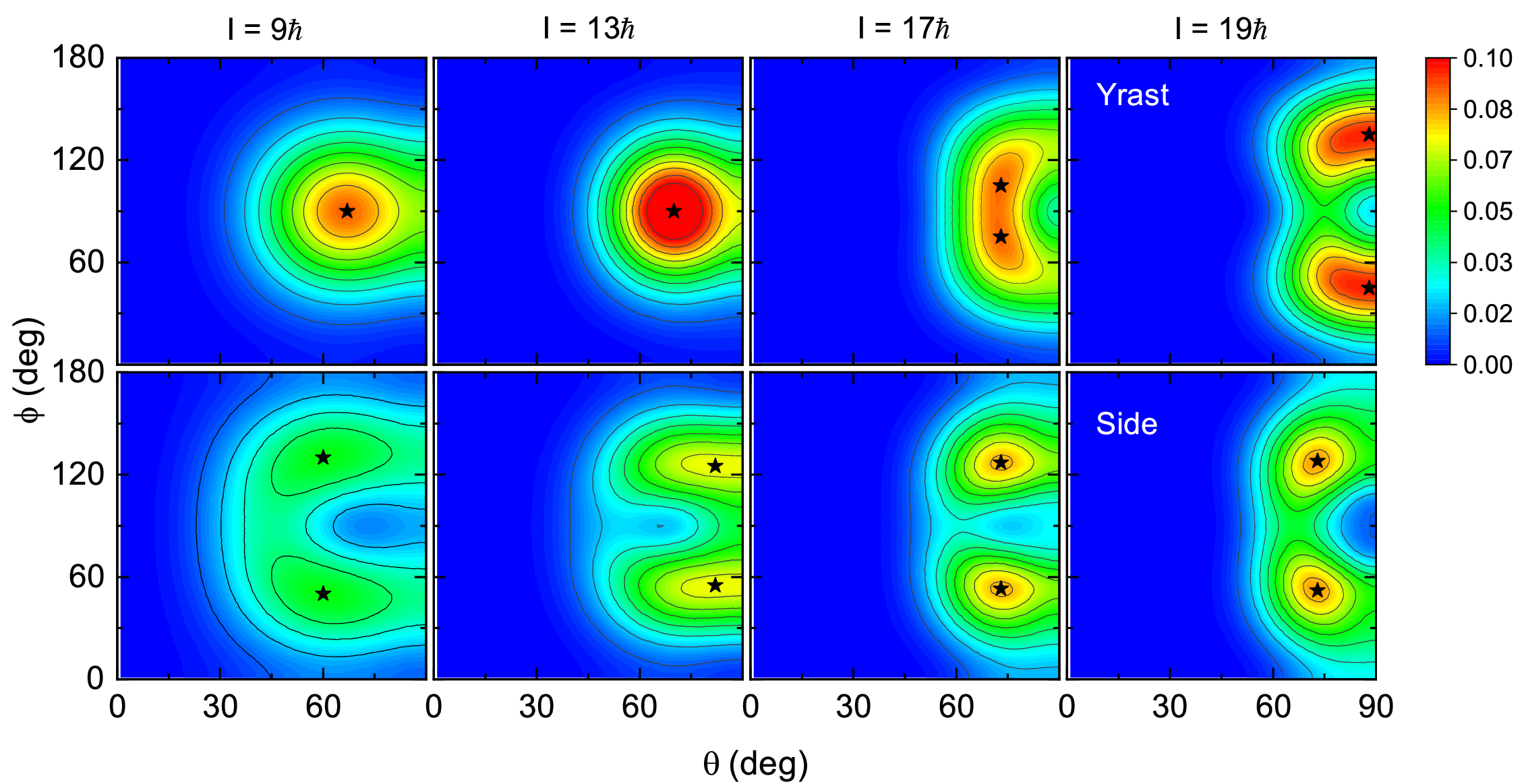}
  \caption{(Color online) The \textit{azimuthal plot}, i.e., the probability distribution profiles for the orientation of the angular momentum on the intrinsic $(\theta,\phi)$ plane for the Yrast band (upper panel) and the Side band (lower panel) at $I = 9\hbar$, $13\hbar$, $17\hbar$, and $19\hbar$.
  The black stars represent the positions of local maxima.}
  \label{Fig:Aplot}
\end{figure}

To determine the spin region in which static chirality is realized in $^{128}$Cs, the \textit{azimuthal plots} extracted from the ReCD many-body wave functions at selected spins are presented in Fig.~\ref{Fig:Aplot}. 
These plots provide a direct visualization of the orientation probability distribution of the total angular momentum in the intrinsic frame and thereby allow us to distinguish among planar rotation, chiral vibration, and static chirality.

At $I=9\hbar$, the \textit{azimuthal plot} of the Yrast state exhibits a single maximum at $(\theta,\phi)=(67^\circ,90^\circ)$. 
Since $\phi=90^\circ$ corresponds to the intrinsic long-short ($l$-$s$) plane, the total angular momentum is predominantly confined to this plane, indicating a planar angular-momentum geometry. 
As discussed in Ref.~\cite{Chen:2017cwl}, such a distribution is characteristic of the zero-phonon state of chiral vibration. 
In contrast, the Side state exhibits two maxima at approximately $(60^\circ,50^\circ)$ and $(60^\circ,130^\circ)$, together with a node around $(73^\circ,90^\circ)$. 
This nodal structure is characteristic of the corresponding one-phonon excitation~\cite{Chen:2017cwl}. 
The complementary zero- and one-phonon patterns of the Yrast and Side states therefore provide a clear signature of chiral vibration at $I=9\hbar$. 
A similar angular-momentum geometry is found at $I=13\hbar$. 
The \textit{azimuthal plots} for $I=10\hbar$, $11\hbar$, $12\hbar$, $14\hbar$, $15\hbar$, and $16\hbar$ exhibit qualitatively similar patterns and are therefore not shown. 
These results demonstrate that the positive-parity doublet bands are dominated by chiral vibration for $I<17\hbar$.

A qualitatively different angular-momentum geometry emerges at $I=17\hbar$. 
For the Yrast state, two probability maxima appear at approximately $(\theta,\phi)=(73^\circ,75^\circ)$ and $(73^\circ,105^\circ)$, whereas the Side state exhibits two maxima around $(73^\circ,53^\circ)$ and $(73^\circ,127^\circ)$. 
In both cases, the two maxima are located symmetrically on opposite sides of the $l$-$s$ plane and correspond to two aplanar orientations of the total angular momentum with opposite handedness. 
The emergence of these two handed configurations in both partner states provides a direct microscopic signature for the realization of static chirality at $I=17\hbar$. 
Although the locations of the probability maxima differ somewhat between the Yrast and Side states, both exhibit the characteristic left- and right-handed angular-momentum geometries.

Interestingly, the static chiral geometry does not persist toward higher spins.
At $I=19\hbar$, two probability maxima remain visible for both the Yrast and Side states. 
For the Yrast state, however, they move close to $\theta=90^\circ$, indicating that the total angular momentum becomes essentially confined to the intrinsic intermediate-short ($i$-$s$) plane. 
The corresponding distribution therefore represents planar rotation rather than an aplanar chiral geometry. 
The same qualitative features are found for the $I=18\hbar$ and $20\hbar$ states. 
These results demonstrate that the static chirality developed around $I=17\hbar$ disappears at higher spins.

The \textit{azimuthal plots} therefore reveal a distinct evolution of the rotational mode in $^{128}$Cs: from chiral vibration at $I<17\hbar$, through static chirality at $I=17\hbar$, to $i$-$s$ planar rotation at $I>17\hbar$ (mainly for Yrast states).
The calculated azimuthal distributions thus indicate that static chiral geometry is confined to a narrow spin region around $I = 17\hbar$.
This microscopic picture is consistent with the qualitative similarities in the spectroscopic observables of the two partner bands around $I=17\hbar$.

\section{Conclusion}
In summary, we have revisited the chiral nature of the positive-parity doublet bands in $^{128}$Cs within the microscopic ReCD framework. 
Starting from self-consistent relativistic DFT calculations and incorporating quasiparticle excitations, three-dimensional angular-momentum projection, and configuration mixing, the energy spectra, $B(E2)$ and $B(M1)$ transition probabilities, $g$ factors, and spectroscopic quadrupole moments have been investigated without introducing additional parameters adjusted to the spectroscopic data of $^{128}$Cs.
Overall, the ReCD calculations provide a satisfactory description of the available experimental data, including the near degeneracy of the positive-parity doublet bands and the characteristic behavior of their electromagnetic transition probabilities. 
A combined examination of the calculated spectroscopic observables and the available experimental data reveals, however, that the similarity between the two partner bands is actually spin dependent. 
In particular, $I=17\hbar$ stands out as the spin at which all the spectroscopic observables of the two bands simultaneously exhibit their most consistent similarity.
More direct insight is obtained from the underlying angular-momentum geometry. 
The \textit{azimuthal plots} reveal characteristic zero- and one-phonon patterns for the Yrast and Side bands at $I < 17\hbar$, identifying this region with chiral vibration. 
At $I = 17\hbar$, two aplanar angular-momentum orientations with opposite handedness emerge for both partner states, providing a direct microscopic signature of static chirality. 
With further increasing spin, however, the angular momentum of the Yrast band evolves toward the intrinsic intermediate-short plane, signaling the disappearance of the static chiral geometry and the development toward planar rotation.
The present results therefore provide a spin-dependent evolution from chiral vibration through static chirality toward planar rotation in $^{128}$Cs. 
Within the present ReCD calculations, static chiral geometry is found only in a narrow spin region around $I=17\hbar$. 
This finding calls for a more nuanced interpretation of $^{128}$Cs, despite its long-standing status as a prominent candidate for nuclear chirality.

\begin{acknowledgments}
This work has been supported in part by the National Natural Science Foundation of China under Grant No. 12105004, the Beijing Natural Science Foundation under Grant No. 1242019, the State Key Laboratory of Dark Matter Physics, and the Fundamental Research Funds for the Central Universities.
\end{acknowledgments}

\section*{DATA AVAILABILITY}
The data that support the findings of this article are not publicly available. 
The data are available from the authors upon reasonable request.


\begin{thebibliography}{67}%
\makeatletter
\providecommand \@ifxundefined [1]{%
 \@ifx{#1\undefined}
}%
\providecommand \@ifnum [1]{%
 \ifnum #1\expandafter \@firstoftwo
 \else \expandafter \@secondoftwo
 \fi
}%
\providecommand \@ifx [1]{%
 \ifx #1\expandafter \@firstoftwo
 \else \expandafter \@secondoftwo
 \fi
}%
\providecommand \natexlab [1]{#1}%
\providecommand \enquote  [1]{``#1''}%
\providecommand \bibnamefont  [1]{#1}%
\providecommand \bibfnamefont [1]{#1}%
\providecommand \citenamefont [1]{#1}%
\providecommand \href@noop [0]{\@secondoftwo}%
\providecommand \href [0]{\begingroup \@sanitize@url \@href}%
\providecommand \@href[1]{\@@startlink{#1}\@@href}%
\providecommand \@@href[1]{\endgroup#1\@@endlink}%
\providecommand \@sanitize@url [0]{\catcode `\\12\catcode `\$12\catcode
  `\&12\catcode `\#12\catcode `\^12\catcode `\_12\catcode `\%12\relax}%
\providecommand \@@startlink[1]{}%
\providecommand \@@endlink[0]{}%
\providecommand \url  [0]{\begingroup\@sanitize@url \@url }%
\providecommand \@url [1]{\endgroup\@href {#1}{\urlprefix }}%
\providecommand \urlprefix  [0]{URL }%
\providecommand \Eprint [0]{\href }%
\providecommand \doibase [0]{https://doi.org/}%
\providecommand \selectlanguage [0]{\@gobble}%
\providecommand \bibinfo  [0]{\@secondoftwo}%
\providecommand \bibfield  [0]{\@secondoftwo}%
\providecommand \translation [1]{[#1]}%
\providecommand \BibitemOpen [0]{}%
\providecommand \bibitemStop [0]{}%
\providecommand \bibitemNoStop [0]{.\EOS\space}%
\providecommand \EOS [0]{\spacefactor3000\relax}%
\providecommand \BibitemShut  [1]{\csname bibitem#1\endcsname}%
\let\auto@bib@innerbib\@empty
\bibitem [{\citenamefont {Frauendorf}\ and\ \citenamefont
  {Meng}(1997)}]{Frauendorf:1997mux}%
  \BibitemOpen
  \bibfield  {author} {\bibinfo {author} {\bibfnamefont {S.}~\bibnamefont
  {Frauendorf}}\ and\ \bibinfo {author} {\bibfnamefont {J.}~\bibnamefont
  {Meng}},\ }\href {https://doi.org/10.1016/S0375-9474(97)00004-3} {\bibfield
  {journal} {\bibinfo  {journal} {Nucl. Phys. A}\ }\textbf {\bibinfo {volume}
  {617}},\ \bibinfo {pages} {131} (\bibinfo {year} {1997})}\BibitemShut
  {NoStop}%
\bibitem [{\citenamefont {Wang}\ \emph {et~al.}(2011)\citenamefont {Wang} \emph
  {et~al.}}]{Wang:2011un}%
  \BibitemOpen
  \bibfield  {author} {\bibinfo {author} {\bibfnamefont {S.~Y.}\ \bibnamefont
  {Wang}} \emph {et~al.},\ }\href
  {https://doi.org/10.1016/j.physletb.2011.07.055} {\bibfield  {journal}
  {\bibinfo  {journal} {Phys. Lett. B}\ }\textbf {\bibinfo {volume} {703}},\
  \bibinfo {pages} {40} (\bibinfo {year} {2011})},\ \Eprint
  {https://arxiv.org/abs/1107.5109} {arXiv:1107.5109 [nucl-ex]} \BibitemShut
  {NoStop}%
\bibitem [{\citenamefont {Liu}\ \emph {et~al.}(2016)\citenamefont {Liu} \emph
  {et~al.}}]{Liu:2016ddt}%
  \BibitemOpen
  \bibfield  {author} {\bibinfo {author} {\bibfnamefont {C.}~\bibnamefont
  {Liu}} \emph {et~al.},\ }\href
  {https://doi.org/10.1103/PhysRevLett.116.112501} {\bibfield  {journal}
  {\bibinfo  {journal} {Phys. Rev. Lett.}\ }\textbf {\bibinfo {volume} {116}},\
  \bibinfo {pages} {112501} (\bibinfo {year} {2016})}\BibitemShut {NoStop}%
\bibitem [{\citenamefont {Vaman}\ \emph {et~al.}(2004)\citenamefont {Vaman},
  \citenamefont {Fossan}, \citenamefont {Koike}, \citenamefont {Starosta},
  \citenamefont {Lee},\ and\ \citenamefont {Macchiavelli}}]{Vaman:2004zz}%
  \BibitemOpen
  \bibfield  {author} {\bibinfo {author} {\bibfnamefont {C.}~\bibnamefont
  {Vaman}}, \bibinfo {author} {\bibfnamefont {D.~B.}\ \bibnamefont {Fossan}},
  \bibinfo {author} {\bibfnamefont {T.}~\bibnamefont {Koike}}, \bibinfo
  {author} {\bibfnamefont {K.}~\bibnamefont {Starosta}}, \bibinfo {author}
  {\bibfnamefont {I.~Y.}\ \bibnamefont {Lee}},\ and\ \bibinfo {author}
  {\bibfnamefont {A.~O.}\ \bibnamefont {Macchiavelli}},\ }\href
  {https://doi.org/10.1103/PhysRevLett.92.032501} {\bibfield  {journal}
  {\bibinfo  {journal} {Phys. Rev. Lett.}\ }\textbf {\bibinfo {volume} {92}},\
  \bibinfo {pages} {032501} (\bibinfo {year} {2004})}\BibitemShut {NoStop}%
\bibitem [{\citenamefont {Joshi}\ \emph {et~al.}(2004)\citenamefont {Joshi}
  \emph {et~al.}}]{Joshi:2004yyg}%
  \BibitemOpen
  \bibfield  {author} {\bibinfo {author} {\bibfnamefont {P.}~\bibnamefont
  {Joshi}} \emph {et~al.},\ }\href
  {https://doi.org/10.1016/j.physletb.2004.05.066} {\bibfield  {journal}
  {\bibinfo  {journal} {Phys. Lett. B}\ }\textbf {\bibinfo {volume} {595}},\
  \bibinfo {pages} {135} (\bibinfo {year} {2004})}\BibitemShut {NoStop}%
\bibitem [{\citenamefont {Tonev}\ \emph {et~al.}(2014)\citenamefont {Tonev}
  \emph {et~al.}}]{Tonev:2014iha}%
  \BibitemOpen
  \bibfield  {author} {\bibinfo {author} {\bibfnamefont {D.}~\bibnamefont
  {Tonev}} \emph {et~al.},\ }\href
  {https://doi.org/10.1103/PhysRevLett.112.052501} {\bibfield  {journal}
  {\bibinfo  {journal} {Phys. Rev. Lett.}\ }\textbf {\bibinfo {volume} {112}},\
  \bibinfo {pages} {052501} (\bibinfo {year} {2014})}\BibitemShut {NoStop}%
\bibitem [{\citenamefont {Lieder}\ \emph {et~al.}(2014)\citenamefont {Lieder}
  \emph {et~al.}}]{Lieder:2014ria}%
  \BibitemOpen
  \bibfield  {author} {\bibinfo {author} {\bibfnamefont {E.~O.}\ \bibnamefont
  {Lieder}} \emph {et~al.},\ }\href
  {https://doi.org/10.1103/PhysRevLett.112.202502} {\bibfield  {journal}
  {\bibinfo  {journal} {Phys. Rev. Lett.}\ }\textbf {\bibinfo {volume} {112}},\
  \bibinfo {pages} {202502} (\bibinfo {year} {2014})}\BibitemShut {NoStop}%
\bibitem [{\citenamefont {Rather}\ \emph {et~al.}(2014)\citenamefont {Rather}
  \emph {et~al.}}]{Rather:2013zos}%
  \BibitemOpen
  \bibfield  {author} {\bibinfo {author} {\bibfnamefont {N.}~\bibnamefont
  {Rather}} \emph {et~al.},\ }\href
  {https://doi.org/10.1103/PhysRevLett.112.202503} {\bibfield  {journal}
  {\bibinfo  {journal} {Phys. Rev. Lett.}\ }\textbf {\bibinfo {volume} {112}},\
  \bibinfo {pages} {202503} (\bibinfo {year} {2014})},\ \Eprint
  {https://arxiv.org/abs/1310.7731} {arXiv:1310.7731 [nucl-ex]} \BibitemShut
  {NoStop}%
\bibitem [{\citenamefont {Kuti}\ \emph {et~al.}(2014)\citenamefont {Kuti} \emph
  {et~al.}}]{Kuti:2014zla}%
  \BibitemOpen
  \bibfield  {author} {\bibinfo {author} {\bibfnamefont {I.}~\bibnamefont
  {Kuti}} \emph {et~al.},\ }\href
  {https://doi.org/10.1103/PhysRevLett.113.032501} {\bibfield  {journal}
  {\bibinfo  {journal} {Phys. Rev. Lett.}\ }\textbf {\bibinfo {volume} {113}},\
  \bibinfo {pages} {032501} (\bibinfo {year} {2014})},\ \Eprint
  {https://arxiv.org/abs/1407.2769} {arXiv:1407.2769 [nucl-ex]} \BibitemShut
  {NoStop}%
\bibitem [{\citenamefont {Starosta}\ \emph {et~al.}(2001)\citenamefont
  {Starosta} \emph {et~al.}}]{Starosta:2001zz}%
  \BibitemOpen
  \bibfield  {author} {\bibinfo {author} {\bibfnamefont {K.}~\bibnamefont
  {Starosta}} \emph {et~al.},\ }\href
  {https://doi.org/10.1103/PhysRevLett.86.971} {\bibfield  {journal} {\bibinfo
  {journal} {Phys. Rev. Lett.}\ }\textbf {\bibinfo {volume} {86}},\ \bibinfo
  {pages} {971} (\bibinfo {year} {2001})}\BibitemShut {NoStop}%
\bibitem [{\citenamefont {Zhu}\ \emph {et~al.}(2003)\citenamefont {Zhu} \emph
  {et~al.}}]{Zhu:2003yu}%
  \BibitemOpen
  \bibfield  {author} {\bibinfo {author} {\bibfnamefont {S.}~\bibnamefont
  {Zhu}} \emph {et~al.},\ }\href
  {https://doi.org/10.1103/PhysRevLett.91.132501} {\bibfield  {journal}
  {\bibinfo  {journal} {Phys. Rev. Lett.}\ }\textbf {\bibinfo {volume} {91}},\
  \bibinfo {pages} {132501} (\bibinfo {year} {2003})},\ \Eprint
  {https://arxiv.org/abs/nucl-ex/0302029} {arXiv:nucl-ex/0302029} \BibitemShut
  {NoStop}%
\bibitem [{\citenamefont {Tonev}\ \emph {et~al.}(2006)\citenamefont {Tonev}
  \emph {et~al.}}]{Tonev:2006zz}%
  \BibitemOpen
  \bibfield  {author} {\bibinfo {author} {\bibfnamefont {D.}~\bibnamefont
  {Tonev}} \emph {et~al.},\ }\href
  {https://doi.org/10.1103/PhysRevLett.96.052501} {\bibfield  {journal}
  {\bibinfo  {journal} {Phys. Rev. Lett.}\ }\textbf {\bibinfo {volume} {96}},\
  \bibinfo {pages} {052501} (\bibinfo {year} {2006})}\BibitemShut {NoStop}%
\bibitem [{\citenamefont {Petrache}\ \emph {et~al.}(2006)\citenamefont
  {Petrache}, \citenamefont {Hagemann}, \citenamefont {Hamamoto},\ and\
  \citenamefont {Starosta}}]{Petrache:2006zz}%
  \BibitemOpen
  \bibfield  {author} {\bibinfo {author} {\bibfnamefont {C.~M.}\ \bibnamefont
  {Petrache}}, \bibinfo {author} {\bibfnamefont {G.~B.}\ \bibnamefont
  {Hagemann}}, \bibinfo {author} {\bibfnamefont {I.}~\bibnamefont {Hamamoto}},\
  and\ \bibinfo {author} {\bibfnamefont {K.}~\bibnamefont {Starosta}},\ }\href
  {https://doi.org/10.1103/PhysRevLett.96.112502} {\bibfield  {journal}
  {\bibinfo  {journal} {Phys. Rev. Lett.}\ }\textbf {\bibinfo {volume} {96}},\
  \bibinfo {pages} {112502} (\bibinfo {year} {2006})}\BibitemShut {NoStop}%
\bibitem [{\citenamefont {Grodner}\ \emph {et~al.}(2006)\citenamefont {Grodner}
  \emph {et~al.}}]{PhysRevLett.97.172501}%
  \BibitemOpen
  \bibfield  {author} {\bibinfo {author} {\bibfnamefont {E.}~\bibnamefont
  {Grodner}} \emph {et~al.},\ }\href
  {https://doi.org/10.1103/PhysRevLett.97.172501} {\bibfield  {journal}
  {\bibinfo  {journal} {Phys. Rev. Lett.}\ }\textbf {\bibinfo {volume} {97}},\
  \bibinfo {pages} {172501} (\bibinfo {year} {2006})}\BibitemShut {NoStop}%
\bibitem [{\citenamefont {Tonev}\ \emph {et~al.}(2007)\citenamefont {Tonev}
  \emph {et~al.}}]{Tonev:2007zz}%
  \BibitemOpen
  \bibfield  {author} {\bibinfo {author} {\bibfnamefont {D.}~\bibnamefont
  {Tonev}} \emph {et~al.},\ }\href {https://doi.org/10.1103/PhysRevC.76.044313}
  {\bibfield  {journal} {\bibinfo  {journal} {Phys. Rev. C}\ }\textbf {\bibinfo
  {volume} {76}},\ \bibinfo {pages} {044313} (\bibinfo {year}
  {2007})}\BibitemShut {NoStop}%
\bibitem [{\citenamefont {Mukhopadhyay}\ \emph {et~al.}(2007)\citenamefont
  {Mukhopadhyay} \emph {et~al.}}]{Mukhopadhyay:2007zx}%
  \BibitemOpen
  \bibfield  {author} {\bibinfo {author} {\bibfnamefont {S.}~\bibnamefont
  {Mukhopadhyay}} \emph {et~al.},\ }\href
  {https://doi.org/10.1103/PhysRevLett.99.172501} {\bibfield  {journal}
  {\bibinfo  {journal} {Phys. Rev. Lett.}\ }\textbf {\bibinfo {volume} {99}},\
  \bibinfo {pages} {172501} (\bibinfo {year} {2007})},\ \Eprint
  {https://arxiv.org/abs/0708.1493} {arXiv:0708.1493 [nucl-ex]} \BibitemShut
  {NoStop}%
\bibitem [{\citenamefont {Ayangeakaa}\ \emph {et~al.}(2013)\citenamefont
  {Ayangeakaa} \emph {et~al.}}]{Ayangeakaa:2013ck}%
  \BibitemOpen
  \bibfield  {author} {\bibinfo {author} {\bibfnamefont {A.~D.}\ \bibnamefont
  {Ayangeakaa}} \emph {et~al.},\ }\href
  {https://doi.org/10.1103/PhysRevLett.110.172504} {\bibfield  {journal}
  {\bibinfo  {journal} {Phys. Rev. Lett.}\ }\textbf {\bibinfo {volume} {110}},\
  \bibinfo {pages} {172504} (\bibinfo {year} {2013})},\ \Eprint
  {https://arxiv.org/abs/1302.0401} {arXiv:1302.0401 [nucl-ex]} \BibitemShut
  {NoStop}%
\bibitem [{\citenamefont {Petrache}\ \emph {et~al.}(2018)\citenamefont
  {Petrache} \emph {et~al.}}]{Petrache:2018tla}%
  \BibitemOpen
  \bibfield  {author} {\bibinfo {author} {\bibfnamefont {C.~M.}\ \bibnamefont
  {Petrache}} \emph {et~al.},\ }\href
  {https://doi.org/10.1103/PhysRevC.97.041304} {\bibfield  {journal} {\bibinfo
  {journal} {Phys. Rev. C}\ }\textbf {\bibinfo {volume} {97}},\ \bibinfo
  {pages} {041304} (\bibinfo {year} {2018})}\BibitemShut {NoStop}%
\bibitem [{\citenamefont {Lv}\ \emph {et~al.}(2019)\citenamefont {Lv} \emph
  {et~al.}}]{Lv:2019ewz}%
  \BibitemOpen
  \bibfield  {author} {\bibinfo {author} {\bibfnamefont {B.~F.}\ \bibnamefont
  {Lv}} \emph {et~al.},\ }\href {https://doi.org/10.1103/PhysRevC.100.024314}
  {\bibfield  {journal} {\bibinfo  {journal} {Phys. Rev. C}\ }\textbf {\bibinfo
  {volume} {100}},\ \bibinfo {pages} {024314} (\bibinfo {year} {2019})},\
  \bibinfo {note} {[Erratum: Phys.Rev.C 103, 019901 (2021)]},\ \Eprint
  {https://arxiv.org/abs/1907.12809} {arXiv:1907.12809 [nucl-ex]} \BibitemShut
  {NoStop}%
\bibitem [{\citenamefont {Balabanski}\ \emph {et~al.}(2004)\citenamefont
  {Balabanski} \emph {et~al.}}]{Balabanski:2004hs}%
  \BibitemOpen
  \bibfield  {author} {\bibinfo {author} {\bibfnamefont {D.~L.}\ \bibnamefont
  {Balabanski}} \emph {et~al.},\ }\href
  {https://doi.org/10.1103/PhysRevC.70.044305} {\bibfield  {journal} {\bibinfo
  {journal} {Phys. Rev. C}\ }\textbf {\bibinfo {volume} {70}},\ \bibinfo
  {pages} {044305} (\bibinfo {year} {2004})}\BibitemShut {NoStop}%
\bibitem [{\citenamefont {Lawrie}\ \emph {et~al.}(2008)\citenamefont {Lawrie}
  \emph {et~al.}}]{Lawrie:2008zzb}%
  \BibitemOpen
  \bibfield  {author} {\bibinfo {author} {\bibfnamefont {E.~A.}\ \bibnamefont
  {Lawrie}} \emph {et~al.},\ }\href
  {https://doi.org/10.1103/PhysRevC.78.021305} {\bibfield  {journal} {\bibinfo
  {journal} {Phys. Rev. C}\ }\textbf {\bibinfo {volume} {78}},\ \bibinfo
  {pages} {021305} (\bibinfo {year} {2008})}\BibitemShut {NoStop}%
\bibitem [{\citenamefont {Koike}\ \emph {et~al.}(2004)\citenamefont {Koike},
  \citenamefont {Starosta},\ and\ \citenamefont {Hamamoto}}]{Koike:2004zz}%
  \BibitemOpen
  \bibfield  {author} {\bibinfo {author} {\bibfnamefont {T.}~\bibnamefont
  {Koike}}, \bibinfo {author} {\bibfnamefont {K.}~\bibnamefont {Starosta}},\
  and\ \bibinfo {author} {\bibfnamefont {I.}~\bibnamefont {Hamamoto}},\ }\href
  {https://doi.org/10.1103/PhysRevLett.93.172502} {\bibfield  {journal}
  {\bibinfo  {journal} {Phys. Rev. Lett.}\ }\textbf {\bibinfo {volume} {93}},\
  \bibinfo {pages} {172502} (\bibinfo {year} {2004})}\BibitemShut {NoStop}%
\bibitem [{\citenamefont {Zhang}\ \emph {et~al.}(2007)\citenamefont {Zhang},
  \citenamefont {Qi}, \citenamefont {Wang},\ and\ \citenamefont
  {Meng}}]{Zhang:2007ge}%
  \BibitemOpen
  \bibfield  {author} {\bibinfo {author} {\bibfnamefont {S.~Q.}\ \bibnamefont
  {Zhang}}, \bibinfo {author} {\bibfnamefont {B.}~\bibnamefont {Qi}}, \bibinfo
  {author} {\bibfnamefont {S.~Y.}\ \bibnamefont {Wang}},\ and\ \bibinfo
  {author} {\bibfnamefont {J.}~\bibnamefont {Meng}},\ }\href
  {https://doi.org/10.1103/PhysRevC.75.044307} {\bibfield  {journal} {\bibinfo
  {journal} {Phys. Rev. C}\ }\textbf {\bibinfo {volume} {75}},\ \bibinfo
  {pages} {044307} (\bibinfo {year} {2007})},\ \Eprint
  {https://arxiv.org/abs/nucl-th/0703047} {arXiv:nucl-th/0703047} \BibitemShut
  {NoStop}%
\bibitem [{\citenamefont {Qi}\ \emph {et~al.}(2009)\citenamefont {Qi},
  \citenamefont {Zhang}, \citenamefont {Meng}, \citenamefont {Wang},\ and\
  \citenamefont {Frauendorf}}]{Qi:2008zp}%
  \BibitemOpen
  \bibfield  {author} {\bibinfo {author} {\bibfnamefont {B.}~\bibnamefont
  {Qi}}, \bibinfo {author} {\bibfnamefont {S.~Q.}\ \bibnamefont {Zhang}},
  \bibinfo {author} {\bibfnamefont {J.}~\bibnamefont {Meng}}, \bibinfo {author}
  {\bibfnamefont {S.~Y.}\ \bibnamefont {Wang}},\ and\ \bibinfo {author}
  {\bibfnamefont {S.}~\bibnamefont {Frauendorf}},\ }\href
  {https://doi.org/10.1016/j.physletb.2009.02.061} {\bibfield  {journal}
  {\bibinfo  {journal} {Phys. Lett. B}\ }\textbf {\bibinfo {volume} {675}},\
  \bibinfo {pages} {175} (\bibinfo {year} {2009})},\ \Eprint
  {https://arxiv.org/abs/0812.4597} {arXiv:0812.4597 [nucl-th]} \BibitemShut
  {NoStop}%
\bibitem [{\citenamefont {Lawrie}\ and\ \citenamefont
  {Shirinda}(2010)}]{Lawrie:2010zzb}%
  \BibitemOpen
  \bibfield  {author} {\bibinfo {author} {\bibfnamefont {E.~A.}\ \bibnamefont
  {Lawrie}}\ and\ \bibinfo {author} {\bibfnamefont {O.}~\bibnamefont
  {Shirinda}},\ }\href {https://doi.org/10.1016/j.physletb.2010.04.047}
  {\bibfield  {journal} {\bibinfo  {journal} {Phys. Lett. B}\ }\textbf
  {\bibinfo {volume} {689}},\ \bibinfo {pages} {66} (\bibinfo {year}
  {2010})}\BibitemShut {NoStop}%
\bibitem [{\citenamefont {Chen}\ \emph {et~al.}(2018)\citenamefont {Chen},
  \citenamefont {Lv}, \citenamefont {Petrache},\ and\ \citenamefont
  {Meng}}]{Chen:2018sys}%
  \BibitemOpen
  \bibfield  {author} {\bibinfo {author} {\bibfnamefont {Q.~B.}\ \bibnamefont
  {Chen}}, \bibinfo {author} {\bibfnamefont {B.~F.}\ \bibnamefont {Lv}},
  \bibinfo {author} {\bibfnamefont {C.~M.}\ \bibnamefont {Petrache}},\ and\
  \bibinfo {author} {\bibfnamefont {J.}~\bibnamefont {Meng}},\ }\href
  {https://doi.org/10.1016/j.physletb.2018.06.030} {\bibfield  {journal}
  {\bibinfo  {journal} {Phys. Lett. B}\ }\textbf {\bibinfo {volume} {782}},\
  \bibinfo {pages} {744} (\bibinfo {year} {2018})},\ \Eprint
  {https://arxiv.org/abs/1805.06162} {arXiv:1805.06162 [nucl-th]} \BibitemShut
  {NoStop}%
\bibitem [{\citenamefont {Wang}\ \emph
  {et~al.}(2019{\natexlab{a}})\citenamefont {Wang}, \citenamefont {Zhang},
  \citenamefont {Zhao},\ and\ \citenamefont {Meng}}]{Wang:2019ehn}%
  \BibitemOpen
  \bibfield  {author} {\bibinfo {author} {\bibfnamefont {Y.~Y.}\ \bibnamefont
  {Wang}}, \bibinfo {author} {\bibfnamefont {S.~Q.}\ \bibnamefont {Zhang}},
  \bibinfo {author} {\bibfnamefont {P.~W.}\ \bibnamefont {Zhao}},\ and\
  \bibinfo {author} {\bibfnamefont {J.}~\bibnamefont {Meng}},\ }\href
  {https://doi.org/10.1016/j.physletb.2019.04.014} {\bibfield  {journal}
  {\bibinfo  {journal} {Phys. Lett. B}\ }\textbf {\bibinfo {volume} {792}},\
  \bibinfo {pages} {454} (\bibinfo {year} {2019}{\natexlab{a}})},\ \Eprint
  {https://arxiv.org/abs/1902.00191} {arXiv:1902.00191 [nucl-th]} \BibitemShut
  {NoStop}%
\bibitem [{\citenamefont {Budaca}(2019)}]{Budaca:2019dnv}%
  \BibitemOpen
  \bibfield  {author} {\bibinfo {author} {\bibfnamefont {R.}~\bibnamefont
  {Budaca}},\ }\href {https://doi.org/10.1016/j.physletb.2019.134853}
  {\bibfield  {journal} {\bibinfo  {journal} {Phys. Lett. B}\ }\textbf
  {\bibinfo {volume} {797}},\ \bibinfo {pages} {134853} (\bibinfo {year}
  {2019})}\BibitemShut {NoStop}%
\bibitem [{\citenamefont {Dimitrov}\ \emph {et~al.}(2000)\citenamefont
  {Dimitrov}, \citenamefont {Frauendorf},\ and\ \citenamefont
  {Doenau}}]{Dimitrov:2000np}%
  \BibitemOpen
  \bibfield  {author} {\bibinfo {author} {\bibfnamefont {V.~I.}\ \bibnamefont
  {Dimitrov}}, \bibinfo {author} {\bibfnamefont {S.}~\bibnamefont
  {Frauendorf}},\ and\ \bibinfo {author} {\bibfnamefont {F.}~\bibnamefont
  {Doenau}},\ }\href {https://doi.org/10.1103/PhysRevLett.84.5732} {\bibfield
  {journal} {\bibinfo  {journal} {Phys. Rev. Lett.}\ }\textbf {\bibinfo
  {volume} {84}},\ \bibinfo {pages} {5732} (\bibinfo {year} {2000})},\ \Eprint
  {https://arxiv.org/abs/nucl-th/0001038} {arXiv:nucl-th/0001038} \BibitemShut
  {NoStop}%
\bibitem [{\citenamefont {Olbratowski}\ \emph {et~al.}(2004)\citenamefont
  {Olbratowski}, \citenamefont {Dobaczewski},\ and\ \citenamefont
  {Dudek}}]{Olbratowski:2002zb}%
  \BibitemOpen
  \bibfield  {author} {\bibinfo {author} {\bibfnamefont {P.}~\bibnamefont
  {Olbratowski}}, \bibinfo {author} {\bibfnamefont {J.}~\bibnamefont
  {Dobaczewski}},\ and\ \bibinfo {author} {\bibfnamefont {J.}~\bibnamefont
  {Dudek}},\ }\href {https://doi.org/10.1103/PhysRevLett.93.052501} {\bibfield
  {journal} {\bibinfo  {journal} {Phys. Rev. Lett.}\ }\textbf {\bibinfo
  {volume} {93}},\ \bibinfo {pages} {052501} (\bibinfo {year} {2004})},\
  \Eprint {https://arxiv.org/abs/nucl-th/0211075} {arXiv:nucl-th/0211075}
  \BibitemShut {NoStop}%
\bibitem [{\citenamefont {Zhao}(2017)}]{Zhao:2017opp}%
  \BibitemOpen
  \bibfield  {author} {\bibinfo {author} {\bibfnamefont {P.~W.}\ \bibnamefont
  {Zhao}},\ }\href {https://doi.org/10.1016/j.physletb.2017.08.001} {\bibfield
  {journal} {\bibinfo  {journal} {Phys. Lett. B}\ }\textbf {\bibinfo {volume}
  {773}},\ \bibinfo {pages} {1} (\bibinfo {year} {2017})},\ \Eprint
  {https://arxiv.org/abs/1706.06127} {arXiv:1706.06127 [nucl-th]} \BibitemShut
  {NoStop}%
\bibitem [{\citenamefont {Chen}\ \emph {et~al.}(2013)\citenamefont {Chen},
  \citenamefont {Zhang}, \citenamefont {Zhao}, \citenamefont {Jolos},\ and\
  \citenamefont {Meng}}]{Chen:2012vk}%
  \BibitemOpen
  \bibfield  {author} {\bibinfo {author} {\bibfnamefont {Q.~B.}\ \bibnamefont
  {Chen}}, \bibinfo {author} {\bibfnamefont {S.~Q.}\ \bibnamefont {Zhang}},
  \bibinfo {author} {\bibfnamefont {P.~W.}\ \bibnamefont {Zhao}}, \bibinfo
  {author} {\bibfnamefont {R.~V.}\ \bibnamefont {Jolos}},\ and\ \bibinfo
  {author} {\bibfnamefont {J.}~\bibnamefont {Meng}},\ }\href
  {https://doi.org/10.1103/PhysRevC.87.024314} {\bibfield  {journal} {\bibinfo
  {journal} {Phys. Rev. C}\ }\textbf {\bibinfo {volume} {87}},\ \bibinfo
  {pages} {024314} (\bibinfo {year} {2013})},\ \Eprint
  {https://arxiv.org/abs/1212.2313} {arXiv:1212.2313 [nucl-th]} \BibitemShut
  {NoStop}%
\bibitem [{\citenamefont {Ren}\ \emph {et~al.}(2022)\citenamefont {Ren},
  \citenamefont {Zhao},\ and\ \citenamefont {Meng}}]{Ren:2022fau}%
  \BibitemOpen
  \bibfield  {author} {\bibinfo {author} {\bibfnamefont {Z.~X.}\ \bibnamefont
  {Ren}}, \bibinfo {author} {\bibfnamefont {P.~W.}\ \bibnamefont {Zhao}},\ and\
  \bibinfo {author} {\bibfnamefont {J.}~\bibnamefont {Meng}},\ }\href
  {https://doi.org/10.1103/PhysRevC.105.L011301} {\bibfield  {journal}
  {\bibinfo  {journal} {Phys. Rev. C}\ }\textbf {\bibinfo {volume} {105}},\
  \bibinfo {pages} {L011301} (\bibinfo {year} {2022})},\ \Eprint
  {https://arxiv.org/abs/2202.03043} {arXiv:2202.03043 [nucl-th]} \BibitemShut
  {NoStop}%
\bibitem [{\citenamefont {Li}\ \emph {et~al.}(2024)\citenamefont {Li},
  \citenamefont {Zhao},\ and\ \citenamefont {Meng}}]{Li:2024fmh}%
  \BibitemOpen
  \bibfield  {author} {\bibinfo {author} {\bibfnamefont {B.}~\bibnamefont
  {Li}}, \bibinfo {author} {\bibfnamefont {P.}~\bibnamefont {Zhao}},\ and\
  \bibinfo {author} {\bibfnamefont {J.}~\bibnamefont {Meng}},\ }\href
  {https://doi.org/10.1016/j.physletb.2024.138877} {\bibfield  {journal}
  {\bibinfo  {journal} {Phys. Lett. B}\ }\textbf {\bibinfo {volume} {856}},\
  \bibinfo {pages} {138877} (\bibinfo {year} {2024})},\ \Eprint
  {https://arxiv.org/abs/2407.08210} {arXiv:2407.08210 [nucl-th]} \BibitemShut
  {NoStop}%
\bibitem [{\citenamefont {Chen}\ \emph {et~al.}(2017)\citenamefont {Chen},
  \citenamefont {Chen}, \citenamefont {Luo}, \citenamefont {Meng},\ and\
  \citenamefont {Zhang}}]{Chen:2017cwl}%
  \BibitemOpen
  \bibfield  {author} {\bibinfo {author} {\bibfnamefont {F.~Q.}\ \bibnamefont
  {Chen}}, \bibinfo {author} {\bibfnamefont {Q.~B.}\ \bibnamefont {Chen}},
  \bibinfo {author} {\bibfnamefont {Y.~A.}\ \bibnamefont {Luo}}, \bibinfo
  {author} {\bibfnamefont {J.}~\bibnamefont {Meng}},\ and\ \bibinfo {author}
  {\bibfnamefont {S.~Q.}\ \bibnamefont {Zhang}},\ }\href
  {https://doi.org/10.1103/PhysRevC.96.051303} {\bibfield  {journal} {\bibinfo
  {journal} {Phys. Rev. C}\ }\textbf {\bibinfo {volume} {96}},\ \bibinfo
  {pages} {051303} (\bibinfo {year} {2017})},\ \Eprint
  {https://arxiv.org/abs/1708.07282} {arXiv:1708.07282 [nucl-th]} \BibitemShut
  {NoStop}%
\bibitem [{\citenamefont {Wang}\ \emph
  {et~al.}(2019{\natexlab{b}})\citenamefont {Wang}, \citenamefont {Chen},
  \citenamefont {Zhao}, \citenamefont {Zhang},\ and\ \citenamefont
  {Meng}}]{Wang:2019inb}%
  \BibitemOpen
  \bibfield  {author} {\bibinfo {author} {\bibfnamefont {Y.~K.}\ \bibnamefont
  {Wang}}, \bibinfo {author} {\bibfnamefont {F.~Q.}\ \bibnamefont {Chen}},
  \bibinfo {author} {\bibfnamefont {P.~W.}\ \bibnamefont {Zhao}}, \bibinfo
  {author} {\bibfnamefont {S.~Q.}\ \bibnamefont {Zhang}},\ and\ \bibinfo
  {author} {\bibfnamefont {J.}~\bibnamefont {Meng}},\ }\href
  {https://doi.org/10.1103/PhysRevC.99.054303} {\bibfield  {journal} {\bibinfo
  {journal} {Phys. Rev. C}\ }\textbf {\bibinfo {volume} {99}},\ \bibinfo
  {pages} {054303} (\bibinfo {year} {2019}{\natexlab{b}})},\ \Eprint
  {https://arxiv.org/abs/1903.03497} {arXiv:1903.03497 [nucl-th]} \BibitemShut
  {NoStop}%
\bibitem [{\citenamefont {Wang}\ \emph
  {et~al.}(2024{\natexlab{a}})\citenamefont {Wang}, \citenamefont {Zhao},\ and\
  \citenamefont {Meng}}]{Wang:2023qll}%
  \BibitemOpen
  \bibfield  {author} {\bibinfo {author} {\bibfnamefont {Y.}~\bibnamefont
  {Wang}}, \bibinfo {author} {\bibfnamefont {P.}~\bibnamefont {Zhao}},\ and\
  \bibinfo {author} {\bibfnamefont {J.}~\bibnamefont {Meng}},\ }\href
  {https://doi.org/10.1016/j.physletb.2023.138346} {\bibfield  {journal}
  {\bibinfo  {journal} {Phys. Lett. B}\ }\textbf {\bibinfo {volume} {848}},\
  \bibinfo {pages} {138346} (\bibinfo {year} {2024}{\natexlab{a}})},\ \Eprint
  {https://arxiv.org/abs/2310.17899} {arXiv:2310.17899 [nucl-th]} \BibitemShut
  {NoStop}%
\bibitem [{\citenamefont {Meng}\ and\ \citenamefont
  {Zhang}(2010)}]{Meng:2010kj}%
  \BibitemOpen
  \bibfield  {author} {\bibinfo {author} {\bibfnamefont {J.}~\bibnamefont
  {Meng}}\ and\ \bibinfo {author} {\bibfnamefont {S.~Q.}\ \bibnamefont
  {Zhang}},\ }\href {https://doi.org/10.1088/0954-3899/37/6/064025} {\bibfield
  {journal} {\bibinfo  {journal} {J. Phys. G}\ }\textbf {\bibinfo {volume}
  {37}},\ \bibinfo {pages} {064025} (\bibinfo {year} {2010})},\ \Eprint
  {https://arxiv.org/abs/1002.0907} {arXiv:1002.0907 [nucl-th]} \BibitemShut
  {NoStop}%
\bibitem [{\citenamefont {Meng}\ and\ \citenamefont
  {Zhao}(2016)}]{Meng:2016jai}%
  \BibitemOpen
  \bibfield  {author} {\bibinfo {author} {\bibfnamefont {J.}~\bibnamefont
  {Meng}}\ and\ \bibinfo {author} {\bibfnamefont {P.}~\bibnamefont {Zhao}},\
  }\href {https://doi.org/10.1088/0031-8949/91/5/053008} {\bibfield  {journal}
  {\bibinfo  {journal} {Phys. Scripta}\ }\textbf {\bibinfo {volume} {91}},\
  \bibinfo {pages} {053008} (\bibinfo {year} {2016})},\ \Eprint
  {https://arxiv.org/abs/1604.02213} {arXiv:1604.02213 [nucl-th]} \BibitemShut
  {NoStop}%
\bibitem [{\citenamefont {Raduta}(2016)}]{Raduta:2016hjl}%
  \BibitemOpen
  \bibfield  {author} {\bibinfo {author} {\bibfnamefont {A.~A.}\ \bibnamefont
  {Raduta}},\ }\href {https://doi.org/10.1016/j.ppnp.2016.05.002} {\bibfield
  {journal} {\bibinfo  {journal} {Prog. Part. Nucl. Phys.}\ }\textbf {\bibinfo
  {volume} {90}},\ \bibinfo {pages} {241} (\bibinfo {year} {2016})},\ \Eprint
  {https://arxiv.org/abs/1705.08220} {arXiv:1705.08220 [nucl-th]} \BibitemShut
  {NoStop}%
\bibitem [{\citenamefont {Chen}\ and\ \citenamefont
  {Meng}(2020)}]{Chen:2020zsy}%
  \BibitemOpen
  \bibfield  {author} {\bibinfo {author} {\bibfnamefont {Q.~B.}\ \bibnamefont
  {Chen}}\ and\ \bibinfo {author} {\bibfnamefont {J.}~\bibnamefont {Meng}},\
  }\href {https://doi.org/10.1080/10619127.2019.1676119} {\bibfield  {journal}
  {\bibinfo  {journal} {Nucl. Phys. News}\ }\textbf {\bibinfo {volume} {30}},\
  \bibinfo {pages} {11} (\bibinfo {year} {2020})}\BibitemShut {NoStop}%
\bibitem [{\citenamefont {Xiong}\ and\ \citenamefont
  {Wang}(2019)}]{Xiong:2018tgf}%
  \BibitemOpen
  \bibfield  {author} {\bibinfo {author} {\bibfnamefont {B.}~\bibnamefont
  {Xiong}}\ and\ \bibinfo {author} {\bibfnamefont {Y.}~\bibnamefont {Wang}},\
  }\href {https://doi.org/10.1016/j.adt.2018.05.002} {\bibfield  {journal}
  {\bibinfo  {journal} {Atom. Data Nucl. Data Tabl.}\ }\textbf {\bibinfo
  {volume} {125}},\ \bibinfo {pages} {193} (\bibinfo {year} {2019})},\ \Eprint
  {https://arxiv.org/abs/1804.04437} {arXiv:1804.04437 [nucl-th]} \BibitemShut
  {NoStop}%
\bibitem [{\citenamefont {Grodner}\ \emph {et~al.}(2018)\citenamefont {Grodner}
  \emph {et~al.}}]{Grodner:2018jgx}%
  \BibitemOpen
  \bibfield  {author} {\bibinfo {author} {\bibfnamefont {E.}~\bibnamefont
  {Grodner}} \emph {et~al.},\ }\href
  {https://doi.org/10.1103/PhysRevLett.120.022502} {\bibfield  {journal}
  {\bibinfo  {journal} {Phys. Rev. Lett.}\ }\textbf {\bibinfo {volume} {120}},\
  \bibinfo {pages} {022502} (\bibinfo {year} {2018})}\BibitemShut {NoStop}%
\bibitem [{\citenamefont {Grodner}\ \emph {et~al.}(2022)\citenamefont {Grodner}
  \emph {et~al.}}]{Grodner:2022cnf}%
  \BibitemOpen
  \bibfield  {author} {\bibinfo {author} {\bibfnamefont {E.}~\bibnamefont
  {Grodner}} \emph {et~al.},\ }\href
  {https://doi.org/10.1103/PhysRevC.106.014318} {\bibfield  {journal} {\bibinfo
   {journal} {Phys. Rev. C}\ }\textbf {\bibinfo {volume} {106}},\ \bibinfo
  {pages} {014318} (\bibinfo {year} {2022})}\BibitemShut {NoStop}%
\bibitem [{\citenamefont {Siwach}\ \emph {et~al.}(2021)\citenamefont {Siwach},
  \citenamefont {Arumugam}, \citenamefont {Ferreira},\ and\ \citenamefont
  {Maglione}}]{Siwach:2021xgj}%
  \BibitemOpen
  \bibfield  {author} {\bibinfo {author} {\bibfnamefont {P.}~\bibnamefont
  {Siwach}}, \bibinfo {author} {\bibfnamefont {P.}~\bibnamefont {Arumugam}},
  \bibinfo {author} {\bibfnamefont {L.~S.}\ \bibnamefont {Ferreira}},\ and\
  \bibinfo {author} {\bibfnamefont {E.}~\bibnamefont {Maglione}},\ }\href
  {https://doi.org/10.1103/PhysRevC.103.024327} {\bibfield  {journal} {\bibinfo
   {journal} {Phys. Rev. C}\ }\textbf {\bibinfo {volume} {103}},\ \bibinfo
  {pages} {024327} (\bibinfo {year} {2021})}\BibitemShut {NoStop}%
\bibitem [{\citenamefont {Meng}(2016)}]{meng2016relativistic}%
  \BibitemOpen
  \bibfield  {author} {\bibinfo {author} {\bibfnamefont {J.}~\bibnamefont
  {Meng}},\ }\href {https://books.google.com/books?id=ImDFCwAAQBAJ} {\emph
  {\bibinfo {title} {Relativistic Density Functional for Nuclear Structure}}},\
  International Review of Nuclear Physics\ (\bibinfo  {publisher} {World
  Scientific Publishing Company},\ \bibinfo {year} {2016})\BibitemShut
  {NoStop}%
\bibitem [{\citenamefont {Caurier}\ \emph {et~al.}(2005)\citenamefont
  {Caurier}, \citenamefont {Martinez-Pinedo}, \citenamefont {Nowacki},
  \citenamefont {Poves},\ and\ \citenamefont {Zuker}}]{Caurier:2004gf}%
  \BibitemOpen
  \bibfield  {author} {\bibinfo {author} {\bibfnamefont {E.}~\bibnamefont
  {Caurier}}, \bibinfo {author} {\bibfnamefont {G.}~\bibnamefont
  {Martinez-Pinedo}}, \bibinfo {author} {\bibfnamefont {F.}~\bibnamefont
  {Nowacki}}, \bibinfo {author} {\bibfnamefont {A.}~\bibnamefont {Poves}},\
  and\ \bibinfo {author} {\bibfnamefont {A.~P.}\ \bibnamefont {Zuker}},\ }\href
  {https://doi.org/10.1103/RevModPhys.77.427} {\bibfield  {journal} {\bibinfo
  {journal} {Rev. Mod. Phys.}\ }\textbf {\bibinfo {volume} {77}},\ \bibinfo
  {pages} {427} (\bibinfo {year} {2005})},\ \Eprint
  {https://arxiv.org/abs/nucl-th/0402046} {arXiv:nucl-th/0402046} \BibitemShut
  {NoStop}%
\bibitem [{\citenamefont {Qu}\ \emph {et~al.}(2026)\citenamefont {Qu},
  \citenamefont {Li},\ and\ \citenamefont {Wang}}]{Qu:2026vsb}%
  \BibitemOpen
  \bibfield  {author} {\bibinfo {author} {\bibfnamefont {T.}~\bibnamefont
  {Qu}}, \bibinfo {author} {\bibfnamefont {B.}~\bibnamefont {Li}},\ and\
  \bibinfo {author} {\bibfnamefont {Y.}~\bibnamefont {Wang}},\ }\href@noop {}
  {\  (\bibinfo {year} {2026})},\ \Eprint {https://arxiv.org/abs/2607.08478}
  {arXiv:2607.08478 [nucl-th]} \BibitemShut {NoStop}%
\bibitem [{\citenamefont {Zhao}\ \emph {et~al.}(2016)\citenamefont {Zhao},
  \citenamefont {Ring},\ and\ \citenamefont {Meng}}]{Zhao:2016umr}%
  \BibitemOpen
  \bibfield  {author} {\bibinfo {author} {\bibfnamefont {P.~W.}\ \bibnamefont
  {Zhao}}, \bibinfo {author} {\bibfnamefont {P.}~\bibnamefont {Ring}},\ and\
  \bibinfo {author} {\bibfnamefont {J.}~\bibnamefont {Meng}},\ }\href
  {https://doi.org/10.1103/PhysRevC.94.041301} {\bibfield  {journal} {\bibinfo
  {journal} {Phys. Rev. C}\ }\textbf {\bibinfo {volume} {94}},\ \bibinfo
  {pages} {041301} (\bibinfo {year} {2016})},\ \Eprint
  {https://arxiv.org/abs/1607.04241} {arXiv:1607.04241 [nucl-th]} \BibitemShut
  {NoStop}%
\bibitem [{\citenamefont {Wang}\ \emph {et~al.}(2022)\citenamefont {Wang},
  \citenamefont {Zhao},\ and\ \citenamefont {Meng}}]{Wang:2022wxl}%
  \BibitemOpen
  \bibfield  {author} {\bibinfo {author} {\bibfnamefont {Y.~K.}\ \bibnamefont
  {Wang}}, \bibinfo {author} {\bibfnamefont {P.~W.}\ \bibnamefont {Zhao}},\
  and\ \bibinfo {author} {\bibfnamefont {J.}~\bibnamefont {Meng}},\ }\href
  {https://doi.org/10.1103/PhysRevC.105.054311} {\bibfield  {journal} {\bibinfo
   {journal} {Phys. Rev. C}\ }\textbf {\bibinfo {volume} {105}},\ \bibinfo
  {pages} {054311} (\bibinfo {year} {2022})},\ \Eprint
  {https://arxiv.org/abs/2204.03866} {arXiv:2204.03866 [nucl-th]} \BibitemShut
  {NoStop}%
\bibitem [{\citenamefont {Qu}\ \emph {et~al.}(2025)\citenamefont {Qu},
  \citenamefont {Wang},\ and\ \citenamefont {Zhao}}]{Qu:2025qvy}%
  \BibitemOpen
  \bibfield  {author} {\bibinfo {author} {\bibfnamefont {T.}~\bibnamefont
  {Qu}}, \bibinfo {author} {\bibfnamefont {Y.~K.}\ \bibnamefont {Wang}},\ and\
  \bibinfo {author} {\bibfnamefont {P.~W.}\ \bibnamefont {Zhao}},\ }\href
  {https://doi.org/10.1103/PhysRevC.111.064309} {\bibfield  {journal} {\bibinfo
   {journal} {Phys. Rev. C}\ }\textbf {\bibinfo {volume} {111}},\ \bibinfo
  {pages} {064309} (\bibinfo {year} {2025})}\BibitemShut {NoStop}%
\bibitem [{\citenamefont {Wang}\ \emph
  {et~al.}(2024{\natexlab{b}})\citenamefont {Wang}, \citenamefont {Zhao},\ and\
  \citenamefont {Meng}}]{Wang:2024zkl}%
  \BibitemOpen
  \bibfield  {author} {\bibinfo {author} {\bibfnamefont {Y.~K.}\ \bibnamefont
  {Wang}}, \bibinfo {author} {\bibfnamefont {P.~W.}\ \bibnamefont {Zhao}},\
  and\ \bibinfo {author} {\bibfnamefont {J.}~\bibnamefont {Meng}},\ }\href
  {https://doi.org/10.1016/j.physletb.2024.138796} {\bibfield  {journal}
  {\bibinfo  {journal} {Phys. Lett. B}\ }\textbf {\bibinfo {volume} {855}},\
  \bibinfo {pages} {138796} (\bibinfo {year} {2024}{\natexlab{b}})},\ \Eprint
  {https://arxiv.org/abs/2403.06455} {arXiv:2403.06455 [nucl-th]} \BibitemShut
  {NoStop}%
\bibitem [{\citenamefont {Wang}\ \emph
  {et~al.}(2024{\natexlab{c}})\citenamefont {Wang}, \citenamefont {Zhao},\ and\
  \citenamefont {Meng}}]{Wang:2023hkc}%
  \BibitemOpen
  \bibfield  {author} {\bibinfo {author} {\bibfnamefont {Y.}~\bibnamefont
  {Wang}}, \bibinfo {author} {\bibfnamefont {P.}~\bibnamefont {Zhao}},\ and\
  \bibinfo {author} {\bibfnamefont {J.}~\bibnamefont {Meng}},\ }\href
  {https://doi.org/10.1016/j.scib.2024.04.071} {\bibfield  {journal} {\bibinfo
  {journal} {Sci. Bull.}\ }\textbf {\bibinfo {volume} {69}},\ \bibinfo {pages}
  {2017} (\bibinfo {year} {2024}{\natexlab{c}})},\ \Eprint
  {https://arxiv.org/abs/2304.12009} {arXiv:2304.12009 [nucl-th]} \BibitemShut
  {NoStop}%
\bibitem [{\citenamefont {Wang}\ \emph {et~al.}(2026)\citenamefont {Wang},
  \citenamefont {Yang},\ and\ \citenamefont {Zhao}}]{Wang:2026kje}%
  \BibitemOpen
  \bibfield  {author} {\bibinfo {author} {\bibfnamefont {Y.~K.}\ \bibnamefont
  {Wang}}, \bibinfo {author} {\bibfnamefont {Y.~L.}\ \bibnamefont {Yang}},\
  and\ \bibinfo {author} {\bibfnamefont {P.~W.}\ \bibnamefont {Zhao}},\ }\href
  {https://doi.org/10.1103/xd9n-4gwf} {\bibfield  {journal} {\bibinfo
  {journal} {Phys. Rev. C}\ }\textbf {\bibinfo {volume} {114}},\ \bibinfo
  {pages} {L011301} (\bibinfo {year} {2026})}\BibitemShut {NoStop}%
\bibitem [{\citenamefont {Ring}\ and\ \citenamefont {Schuck}(2004)}]{Ring2004}%
  \BibitemOpen
  \bibfield  {author} {\bibinfo {author} {\bibfnamefont {P.}~\bibnamefont
  {Ring}}\ and\ \bibinfo {author} {\bibfnamefont {P.}~\bibnamefont {Schuck}},\
  }\href@noop {} {\emph {\bibinfo {title} {The nuclear many-body problem}}}\
  (\bibinfo  {publisher} {Springer Science \& Business Media, New York},\
  \bibinfo {year} {2004})\BibitemShut {NoStop}%
\bibitem [{\citenamefont {Hu}\ \emph {et~al.}(2014)\citenamefont {Hu},
  \citenamefont {Gao},\ and\ \citenamefont {Chen}}]{Hu:2013dha}%
  \BibitemOpen
  \bibfield  {author} {\bibinfo {author} {\bibfnamefont {Q.-L.}\ \bibnamefont
  {Hu}}, \bibinfo {author} {\bibfnamefont {Z.-C.}\ \bibnamefont {Gao}},\ and\
  \bibinfo {author} {\bibfnamefont {Y.~S.}\ \bibnamefont {Chen}},\ }\href
  {https://doi.org/10.1016/j.physletb.2014.05.045} {\bibfield  {journal}
  {\bibinfo  {journal} {Phys. Lett. B}\ }\textbf {\bibinfo {volume} {734}},\
  \bibinfo {pages} {162} (\bibinfo {year} {2014})},\ \Eprint
  {https://arxiv.org/abs/1307.6905} {arXiv:1307.6905 [nucl-th]} \BibitemShut
  {NoStop}%
\bibitem [{\citenamefont {Carlsson}\ and\ \citenamefont
  {Rotureau}(2021)}]{Carlsson:2020fiv}%
  \BibitemOpen
  \bibfield  {author} {\bibinfo {author} {\bibfnamefont {B.~G.}\ \bibnamefont
  {Carlsson}}\ and\ \bibinfo {author} {\bibfnamefont {J.}~\bibnamefont
  {Rotureau}},\ }\href {https://doi.org/10.1103/PhysRevLett.126.172501}
  {\bibfield  {journal} {\bibinfo  {journal} {Phys. Rev. Lett.}\ }\textbf
  {\bibinfo {volume} {126}},\ \bibinfo {pages} {172501} (\bibinfo {year}
  {2021})},\ \Eprint {https://arxiv.org/abs/2010.08459} {arXiv:2010.08459
  [nucl-th]} \BibitemShut {NoStop}%
\bibitem [{\citenamefont {Zhao}\ \emph {et~al.}(2010)\citenamefont {Zhao},
  \citenamefont {Li}, \citenamefont {Yao},\ and\ \citenamefont
  {Meng}}]{Zhao:2010hi}%
  \BibitemOpen
  \bibfield  {author} {\bibinfo {author} {\bibfnamefont {P.~W.}\ \bibnamefont
  {Zhao}}, \bibinfo {author} {\bibfnamefont {Z.~P.}\ \bibnamefont {Li}},
  \bibinfo {author} {\bibfnamefont {J.~M.}\ \bibnamefont {Yao}},\ and\ \bibinfo
  {author} {\bibfnamefont {J.}~\bibnamefont {Meng}},\ }\href
  {https://doi.org/10.1103/PhysRevC.82.054319} {\bibfield  {journal} {\bibinfo
  {journal} {Phys. Rev. C}\ }\textbf {\bibinfo {volume} {82}},\ \bibinfo
  {pages} {054319} (\bibinfo {year} {2010})},\ \Eprint
  {https://arxiv.org/abs/1002.1789} {arXiv:1002.1789 [nucl-th]} \BibitemShut
  {NoStop}%
\bibitem [{\citenamefont {Tian}\ \emph {et~al.}(2009)\citenamefont {Tian},
  \citenamefont {Ma},\ and\ \citenamefont {Ring}}]{Tian:2009zzh}%
  \BibitemOpen
  \bibfield  {author} {\bibinfo {author} {\bibfnamefont {Y.}~\bibnamefont
  {Tian}}, \bibinfo {author} {\bibfnamefont {Z.~Y.}\ \bibnamefont {Ma}},\ and\
  \bibinfo {author} {\bibfnamefont {P.}~\bibnamefont {Ring}},\ }\href
  {https://doi.org/10.1016/j.physletb.2009.04.067} {\bibfield  {journal}
  {\bibinfo  {journal} {Phys. Lett. B}\ }\textbf {\bibinfo {volume} {676}},\
  \bibinfo {pages} {44} (\bibinfo {year} {2009})},\ \Eprint
  {https://arxiv.org/abs/0908.1844} {arXiv:0908.1844 [nucl-th]} \BibitemShut
  {NoStop}%
\bibitem [{\citenamefont {Zhao}\ \emph {et~al.}(2015)\citenamefont {Zhao},
  \citenamefont {Zhang},\ and\ \citenamefont {Meng}}]{Zhao:2015bua}%
  \BibitemOpen
  \bibfield  {author} {\bibinfo {author} {\bibfnamefont {P.~W.}\ \bibnamefont
  {Zhao}}, \bibinfo {author} {\bibfnamefont {S.~Q.}\ \bibnamefont {Zhang}},\
  and\ \bibinfo {author} {\bibfnamefont {J.}~\bibnamefont {Meng}},\ }\href
  {https://doi.org/10.1103/PhysRevC.92.034319} {\bibfield  {journal} {\bibinfo
  {journal} {Phys. Rev. C}\ }\textbf {\bibinfo {volume} {92}},\ \bibinfo
  {pages} {034319} (\bibinfo {year} {2015})},\ \Eprint
  {https://arxiv.org/abs/1506.08340} {arXiv:1506.08340 [nucl-th]} \BibitemShut
  {NoStop}%
\bibitem [{\citenamefont {Wang}\ and\ \citenamefont
  {Meng}(2023)}]{Wang:2022xls}%
  \BibitemOpen
  \bibfield  {author} {\bibinfo {author} {\bibfnamefont {Y.~P.}\ \bibnamefont
  {Wang}}\ and\ \bibinfo {author} {\bibfnamefont {J.}~\bibnamefont {Meng}},\
  }\href {https://doi.org/10.1016/j.physletb.2023.137923} {\bibfield  {journal}
  {\bibinfo  {journal} {Phys. Lett. B}\ }\textbf {\bibinfo {volume} {841}},\
  \bibinfo {pages} {137923} (\bibinfo {year} {2023})},\ \Eprint
  {https://arxiv.org/abs/2212.04146} {arXiv:2212.04146 [nucl-th]} \BibitemShut
  {NoStop}%
\bibitem [{\citenamefont {Guo}\ \emph {et~al.}(2019)\citenamefont {Guo},
  \citenamefont {Sun}, \citenamefont {Li}, \citenamefont {Yang}, \citenamefont
  {Liu}, \citenamefont {Ru},\ and\ \citenamefont {Chi}}]{Guo:2019gtc}%
  \BibitemOpen
  \bibfield  {author} {\bibinfo {author} {\bibfnamefont {R.}~\bibnamefont
  {Guo}}, \bibinfo {author} {\bibfnamefont {W.-J.}\ \bibnamefont {Sun}},
  \bibinfo {author} {\bibfnamefont {J.}~\bibnamefont {Li}}, \bibinfo {author}
  {\bibfnamefont {D.}~\bibnamefont {Yang}}, \bibinfo {author} {\bibfnamefont
  {Y.}~\bibnamefont {Liu}}, \bibinfo {author} {\bibfnamefont {C.}~\bibnamefont
  {Ru}},\ and\ \bibinfo {author} {\bibfnamefont {J.}~\bibnamefont {Chi}},\
  }\href {https://doi.org/10.1103/PhysRevC.100.034328} {\bibfield  {journal}
  {\bibinfo  {journal} {Phys. Rev. C}\ }\textbf {\bibinfo {volume} {100}},\
  \bibinfo {pages} {034328} (\bibinfo {year} {2019})},\ \Eprint
  {https://arxiv.org/abs/1910.04523} {arXiv:1910.04523 [nucl-th]} \BibitemShut
  {NoStop}%
\bibitem [{\citenamefont {Harting}\ \emph {et~al.}(1981)\citenamefont
  {Harting}, \citenamefont {Weise}, \citenamefont {Toki},\ and\ \citenamefont
  {Richter}}]{Harting:1981cc}%
  \BibitemOpen
  \bibfield  {author} {\bibinfo {author} {\bibfnamefont {A.}~\bibnamefont
  {Harting}}, \bibinfo {author} {\bibfnamefont {W.}~\bibnamefont {Weise}},
  \bibinfo {author} {\bibfnamefont {H.}~\bibnamefont {Toki}},\ and\ \bibinfo
  {author} {\bibfnamefont {A.}~\bibnamefont {Richter}},\ }\href
  {https://doi.org/10.1016/0370-2693(81)90122-2} {\bibfield  {journal}
  {\bibinfo  {journal} {Phys. Lett. B}\ }\textbf {\bibinfo {volume} {104}},\
  \bibinfo {pages} {261} (\bibinfo {year} {1981})}\BibitemShut {NoStop}%
\bibitem [{\citenamefont {Kohno}\ and\ \citenamefont
  {Sprung}(1982)}]{Kohno:1982zz}%
  \BibitemOpen
  \bibfield  {author} {\bibinfo {author} {\bibfnamefont {M.}~\bibnamefont
  {Kohno}}\ and\ \bibinfo {author} {\bibfnamefont {D.~W.~L.}\ \bibnamefont
  {Sprung}},\ }\href {https://doi.org/10.1103/PhysRevC.26.297} {\bibfield
  {journal} {\bibinfo  {journal} {Phys. Rev. C}\ }\textbf {\bibinfo {volume}
  {26}},\ \bibinfo {pages} {297} (\bibinfo {year} {1982})}\BibitemShut
  {NoStop}%
\bibitem [{\citenamefont {Chen}\ \emph {et~al.}(2020)\citenamefont {Chen},
  \citenamefont {Kaiser}, \citenamefont {Mei{\ss}ner},\ and\ \citenamefont
  {Meng}}]{Chen:2020kwc}%
  \BibitemOpen
  \bibfield  {author} {\bibinfo {author} {\bibfnamefont {Q.~B.}\ \bibnamefont
  {Chen}}, \bibinfo {author} {\bibfnamefont {N.}~\bibnamefont {Kaiser}},
  \bibinfo {author} {\bibfnamefont {U.-G.}\ \bibnamefont {Mei{\ss}ner}},\ and\
  \bibinfo {author} {\bibfnamefont {J.}~\bibnamefont {Meng}},\ }\href
  {https://doi.org/10.1016/j.physletb.2020.135568} {\bibfield  {journal}
  {\bibinfo  {journal} {Phys. Lett. B}\ }\textbf {\bibinfo {volume} {807}},\
  \bibinfo {pages} {135568} (\bibinfo {year} {2020})},\ \Eprint
  {https://arxiv.org/abs/2005.03865} {arXiv:2005.03865 [nucl-th]} \BibitemShut
  {NoStop}%
\bibitem [{\citenamefont {Hu}\ and\ \citenamefont {Chen}(2024)}]{Hu:2024vwn}%
  \BibitemOpen
  \bibfield  {author} {\bibinfo {author} {\bibfnamefont {B.}~\bibnamefont
  {Hu}}\ and\ \bibinfo {author} {\bibfnamefont {Q.~B.}\ \bibnamefont {Chen}},\
  }\href {https://doi.org/10.1103/PhysRevC.109.L021302} {\bibfield  {journal}
  {\bibinfo  {journal} {Phys. Rev. C}\ }\textbf {\bibinfo {volume} {109}},\
  \bibinfo {pages} {L021302} (\bibinfo {year} {2024})}\BibitemShut {NoStop}%
\bibitem [{\citenamefont {Chen}(2024)}]{Chen:2024psf}%
  \BibitemOpen
  \bibfield  {author} {\bibinfo {author} {\bibfnamefont {Q.~B.}\ \bibnamefont
  {Chen}},\ }\href {https://doi.org/10.1103/PhysRevC.109.024308} {\bibfield
  {journal} {\bibinfo  {journal} {Phys. Rev. C}\ }\textbf {\bibinfo {volume}
  {109}},\ \bibinfo {pages} {024308} (\bibinfo {year} {2024})}\BibitemShut
  {NoStop}%
\end{thebibliography}
%

\end{document}